\documentclass[referee]{mn2e}

\usepackage[usenames]{color}

\newcommand{\be}{\begin{equation}}
\newcommand{\ee}{\end{equation}}

\usepackage{graphicx}
\usepackage{txfonts}
\title[The GeV to TeV connection in SNR  IC 443]{The GeV to TeV connection in the 
environment of SNR  IC 443 }

\author[Torres, Rodriguez Marrero \& de Cea del Pozo]{
Diego F. Torres$^{1,2}\thanks{E-mail: dtorres@ieec.uab.es}$,
Ana Y. Rodr\'iguez Marrero$^{3}\thanks{E-mail: arodrig@ifca.unican.es}, $ \& Elsa de Cea del Pozo$^{2}\thanks{E-mail: decea@ieec.uab.es}$\\
$^1$Instituci\'o Catalana de Recerca i Estudis Avan\c{c}ats (ICREA)\\
$^2$Institut de Ci\`encies de l'Espai (IEEC-CSIC),
              Campus UAB,  Torre C5, 2a planta,
              08193 Barcelona, Spain\\
$^3$Instituto de F'sica de Cantabria,
Edificio Juan Jord\'a,
Avenida de los Castros s/n, 
39005 Santander,
Cantabria, Spain          
                            }

\begin{document}

\date{}

\pagerange{\pageref{firstpage}--\pageref{lastpage}} \pubyear{2008}

\maketitle

\label{firstpage}

\begin{abstract}

We have recently interpreted the source MAGIC J0616+225 as a result of delayed TeV emission of cosmic-rays diffusing from IC 443 and interacting with a cloud  in the foreground of the remnant. This model was used to make predictions for future observations, especially those to be made with the {\it   Fermi} satellite. 
Just recently, AGILE, {\it   Fermi}, and VERITAS have released new results of their observations of IC 443. In this work, we compare them with the predictions of our model, exploring the GeV to TeV connection in this region of space. We use {\it Fermi} data to consider  the possibility of constraining the cosmic-ray diffusion features of the environment. We analyze the cosmic-ray distributions, their interactions, and a possible detection of the SNR environment in the neutrino channel.

\end{abstract}


\section{Introduction}

{ It is commonly accepted that supernova remnants (SNR) are one of the most probable scenarios of leptonic and hadronic cosmic-ray (CR) acceleration. The particle acceleration mechanism in individual SNRs is usually assumed to be diffusive shock acceleration, which naturally leads to a power-law population of relativistic particles. In the standard version of this mechanism (e.g. Bell 1978), particles are scattered by magnetohydrodynamic waves repeatedly through the shock front. Electrons suffer synchrotron losses, producing the non-thermal emission from radio to X-rays usually seen in shell-type SNRs. The maximum energy achieved depends on the shock speed and age as well as on any competing loss processes. In young SNRs, electrons can easily reach energies in excess of 1 TeV, and they produce X-rays. Non-thermal X-ray emission associated with shock acceleration has been clearly observed in many SNRs. But in order to have an observational confirmation of protons and other nuclei being accelerated, particularly, in order to be able to distinguish this from leptonic emission,  one should try and isolate the multi-messenger effects of the secondary particles produced when the accelerated hadrons interact in nearby molecular clouds through $pp$ collisions. These ideas go back, for instance, to the works by Dogel \& Sharov 1990, Naito \& Takahara 1994; Drury 1994; Sturner et al. 1997; Gaisser et al. 1998; Baring et al. 1999, among others. In fact, as early as 1979, Montmerle suggested that SNRs within OB stellar associations, i.e. star forming regions with plenty of molecular gas, could generate observable $\gamma$-ray sources. A molecular cloud being illuminated by particles that escaped from a nearby SNR could then act as a target for $pp$ interactions, greatly enhancing the $\gamma$-ray emission (see, e.g., the recent works by Gabici et al. 2007, 2009; Casanova et al. 2009; Rodriguez-Marrero et al. 2008). As an spinoff, observing $\gamma$-rays from clouds nearby SNRs, can feedback on our knowledge of the diffusion characteristics of the environment.
As has been emphasized by Aharonian \& Atoyan (1996), the
observed $\gamma$-rays can have a significantly different
spectrum from that expected from the primary particle population at the immediate vicinity of 
source (the SNR shock). 
For instance, a standard diffusion coefficients $\delta\sim
0.3-0.6$ can explain $\gamma$-ray spectra as steep as $\Gamma\sim
2.3-2.6$ in sources with particles accelerated to a power-law
$J_p(E_p)\propto E^{-2}$ if the target that is illuminated by the
$\pi^0$-decays is at sufficient distance from the accelerator. Measuring $\gamma$-ray emission around SNRs would then allow to acquire
knowledge of the diffusion environment in which the CRs propagate, at several kpc from Earth. 
}

Of all SNRs that were found to be positionally coincident with $\gamma$-ray sources in the MeV range in the EGRET era, IC 443 was one of the most appealing for subsequent observations with higher sensitivity instruments { (see the case-by-case study by Torres et al. 2003)}. It was, perhaps with W28, the only case in which the molecular environment { --as mapped for instance with CO observations--} showed a peak in density close by, but separated in sky projection, from the SNR center. This would allow distinguishing, in case the $\gamma$-ray emission observed would be hadronically produced, possible cosmic-ray diffusion effects. Along the last year, several new observations of the IC 443 environment
have been made, and in this work, we consider these in the setting of a theoretical model in which CRs from the SNR IC 443 are diffusing away from it and interacting with clouds nearby.  { This model was originally put forward by Aharonian \& Atoyan (1996), and Torres et al. (2008), referred to as Paper I in this work, studied this model for IC 443 prior to the new wealth of data we can now consider.}

\section{High and very high-energy observations}

\subsection{Earlier EGRET and MAGIC observations}
MAGIC observations towards IC 443 yielded the detection of
J0616+225 nearby, but displaced from the center of the SNR IC 443, with centroid 
located at (RA,DEC)$_{J2000}$=(06$^\mathrm{h}$16$^\mathrm{m}$43$^\mathrm{s}$,
+22$^\circ$31' 48''), $\pm 0.025^\circ_{stat} \pm 0.017^\circ_{sys}$
(Albert et al. 2007). No extension nor any variability was claimed in the $\gamma$-ray data. Albert et al. (2007) showed that 
the MAGIC source is located at the position of a giant cloud in front of the SNR.
A simple power law was fitted to the measured
spectral points:
$
{ \mathrm{d}N_{\gamma}}/ ({\mathrm{d}A \mathrm{d}t \mathrm{d}E})
= (1.0 \pm 0.2_{stat} \pm 0.35_{sys}) \times  10^{-11} 
\left({E}/{\mathrm{0.4 TeV}}\right)^{-3.1 \pm 0.3_{stat} \pm 0.2_{sys}}
 \mathrm{cm}^{-2}\mathrm{s}^{-1}
\mathrm{TeV}^{-1}.$
The integral flux of MAGIC~J0616+225 above 100~GeV is about 6.5\%
of the Crab Nebula. 
The EGRET flux of the source 3EG J0617+2238, which is positionally correlated with the SNR IC 443, is (51.4$\pm$3.5) $\times 10^{-8}$ ph cm$^{-2}$ s$^{-1}$, and it presents a photon spectral index of 2.01$\pm$0.06 (Hartman et al. 1999). The EGRET source was classified as non-variable by Torres et al. (2001) and Nolan et al. (2003).
An independent analysis of GeV photons measured by EGRET resulted
in the source GeV~J0617+2237 (Lamb \& Macomb 1997),
also at the same location of 3EG~J0617+2238, the centroid of which is at the center of the SNR shell.

\subsection{Recent TeV observations} 
Recently, the Very Energetic Radiation Imaging Telescope Array System (VERITAS) presented further observations towards IC 443 (Acciari et al. 2009). Regarding the position of the centroid, it was found to be
at (RA,DEC)$_{J2000}$=(06$^\mathrm{h}$16$^\mathrm{m}$51$^\mathrm{s}$,+22$^\circ$30' 11''), $\pm 0.03^\circ_{stat} \pm 0.08^\circ_{sys}$ thus,  consistent with that of MAGIC. 
Evidence that the very-high-energy (VHE, $E > 100$ GeV) $\gamma$-ray emission is extended was also found.
The extension derived was $0.16^\circ \pm 0.03^\circ_{stat} \pm 0.04^\circ_{sys}$.  
The VHE spectrum is well fit by a power law ($dN/dE = N_0 \times (E/\textrm{TeV})^{-\Gamma}$) with a photon index of $2.99 \pm 0.38_{stat} \pm 0.3_{sys}$ and an integral flux above 300 GeV of $(4.63 \pm 0.90_{stat} \pm 0.93_{sys}) \times 10^{-12}\ \textrm{cm}^{-2}\ \textrm{s}^{-1}$. { Thus, as we will graphically see below, the spectral determination is consistent with the MAGIC measurements, both present a steep slope, with VERITAS finding a slight overall increase in the flux level. }
No variability of the $\gamma$-ray emission was claimed by VERITAS either.

\subsection{Recent GeV observations}
AGILE results on IC 443 has been recently reported too (Tavani et al. 2010). AGILE discovered a distinct pattern of diffuse emission in the energy range 100 MeV--3 GeV coming from the SNR, with a prominent maximum  localized in the Northeastern shell, dislocated (as it was the case with EGRET) with the MAGIC/VERITAS sources. The latter is $\sim$0.4$^o$ apart from the maximum of the AGILE emission (which in turn is also away from the nearby PWN, discussed below). 
Finally, {\it   Fermi} has also recently presented an analysis of its first 11 months of observations towards the region of SNR IC 443 (Abdo et el. 2010). These results enhance, given the better instrument sensitivity, those obtained by AGILE. Thus, we focus on {\it   Fermi} measurements when analyzing GeV results. 
The source was detected in a broad range of energies, from 200 MeV up to 50 GeV,  with a SED that rolls over at about 3 GeV to seemingly match in slope the one that is found at the highest energies; i.e., it can be represented, for instance, with a broken power law with slopes of 1.93 $\pm$ 0.03 and 2.56 $\pm$ 0.11 and with a break at $3.25 \pm 0.6$ GeV. This is one important difference with EGRET data, which SED did not allow to suspect neither that the emission would maintain a hard spectrum up to such tens-of-GeV energies nor the existence of a roll over in the spectrum at the energies found. The flux above 200 MeV resulted to be (28.5$\pm$0.7) $\times 10^{-8}$ ph cm$^{-2}$ s$^{-1}$ what allowed for a very significant detection in {\it   Fermi}.  
The centroid of the emission is consistent with that of EGRET 3EG J0617+2238.

\subsection { Relative localization of sources}

Abdo et al. (2010) report that the centroid of the {\it   Fermi} emission  is displaced more than $5 \times \theta_{68}^{error}$(MAGIC error) from that of MAGIC
(J0610+225), and more than $1.5 \times \theta_{68}^{error}$ (VERITAS error) from that of the
VERITAS source. These numbers are obtained assuming that the systematic and statistical errors in localization add up in quadrature,
and considering the worse error of each of the pairs of measurements ({\it   Fermi}--MAGIC, {\it   Fermi}--VERITAS), which in both cases correspond to the IACTs.  
The significance of the separation greatly improves when a) the best measured position is considered (i.e., the error by {\it   Fermi}), for which both pairs of measurements are about 5$\sigma$ away, and/or b) when statistical errors only are considered for VERITAS (the systematic errors in this latter measurement is about a factor of 3 larger than the statistics and significantly different from all others, but of course, one can not necessarily  assume it to approach the detection in the direction of the {\it   Fermi} source). Thus, albeit  current measurements are not conclusive about energy dependent morphology, they are consistent with it: 
Abdo et al. (2010) report that the centroid of the emission moves (but still not significantly in {\it   Fermi} data: only at $\sim 1.5 \sigma$)
towards that of the VERITAS source as the energy band changes from 1-5 GeV to 5-50 GeV. 
It might be that the angular resolution and/or sensitvity and/or the separation of the real molecular mass distribution on sky projection are not
enough to distinguish the difference when such nearby energy ranges are considered.
New measurements from MAGIC (using the just-obtained stereoscopic capability of the array) could provide continuous coverage from 50 GeV up.

\subsection{A PWN?}

In all energy bands, the centroid of the correspondingly detected sources is inconsistent with the  pulsar wind nebula (and the putative pulsar)
CXOU J061705.3+222127, discovered by Olbert et al. (2001), and lying nearby.
Both the 3EG and the GeV source in the catalogs of Hartman et al. (1999) 
and Lamb \& Macomb (1997), which are co-spatial, are inconsistent with the PWN
location.   Similarly, using the position of the PWN 
and the {\it   Fermi} source one sees that they are
separated by 0.26$^o$,
or about 11$\sigma$ away from the localization of the {\it   Fermi} peak.
Also at higher energies, the $\gamma$-ray emission observed by VERITAS and MAGIC is offset from the location of the PWN by 10-20 arcmin.
This latter fact could be understood in case the PWN is a $\gamma$-ray emitter,  it would be similar to the case of HESS J1825-137 (Aharonian et al. 2006) or HESS J1908+063 (Aharonian et al. 2009), where similar offsets were found, see also Abdo et al. (2010b). The emission could be consistent with a scenario in which the VHE emission arises from inverse Compton scattering off electrons accelerated early in the PWN's life. 
However, if one would assume that the PWN CXOU J061705.3+222127 is producing the emission (note that pulsed radiation from this object has not been found at any frequency), the highest energy TeV-band radiation 
should peak there (it could be extended, but due to losses, the higher the energy, the more peaked towards the PWN the emission will
be) and the GeV radiation should then be unresolved pulsar
emission, it should also peak there and be pulsed (see Bartko \& Bednarek 2008). Then, based on
{\it   Fermi}/VERITAS data we can safely entertain that the GeV and TeV emissions detected do not originate in the PWN, what we explore here further.

\section{The cosmic-ray diffusion model}

As a first approach to the modeling of the GeV to TeV SED, 
Abdo et al. (2010) assumed that a single proton spectrum was directly interacting with the whole molecular mass found in the IC 443 environment (see Torres et al. 2003). Given that the location of sources at different energies change, and that target material for accelerated cosmic-rays are also found to be at different positions, this approach is just a crude approximation to the need of considering cosmic-ray diffusion. This kind of model was introduced in Paper I, and we refer the reader there for details: basically, we computed the spectrum of $\gamma$-rays generated through $\pi^0$-decay at
a source of proton density $n_{p}$ (see e.g., Torres 2004 or Domingo-Santamaria \& Torres 2005 for the formulae we used); solving for the cosmic ray spectrum at each distance of the SNR, neglecting temporal or spatial effects (non-uniform densities) within the
molecular cloud itself.
{
The spectrum of $\gamma$-rays generated through $\pi^0$-decay at a source of proton density $n_{p}$ is 
\be
F_{\gamma}(E_{\gamma})=2\int^{\infty}_{E_{\pi}^{\rm min}}
({F_{\pi}(E_{\pi})}/{\sqrt{E_{\pi}^{2}-m_{\pi}^{2}}}) \;dE_{\pi},
\ee
where  
the minimum pion energy is
$
E_{\pi}^{\rm min}(E_{\gamma})=E_{\gamma}+ {m_{\pi}^{2}}/{4E_{\gamma}}, 
$
and
\be
F_{\pi}(E_{\pi})=4\pi n_{p}\int^{E^{\rm max}_{p}}_{E^{\rm
min}_{p}} J_p(E) ({d\sigma_{\pi}(E_{\pi},\;E_{p})}/{dE_{\pi}})
\;dE_{p}.
\ee
Here, $d\sigma_{\pi}(E_{\pi},\;E_{p})/dE_{\pi}$ is the
differential cross-section for the production of $\pi^0$-mesons of
energy $E_{\pi}$ by a proton of energy $E_{p}$ in a $pp$
collision. We analyze below the influence upon the results of different parameterizations of this cross section. We include cosmic-rays and target nuclei heavier than the proton throughout this paper. We adopt here the approximation in which these can be accounted for by multiplying a nuclear factor (adopted as 1.5) to the total flux, without changing the cosmic-ray proton spectrum (Gaisser \& Schaefer 1992). }

{ We have assumed a uniform cosmic-ray and gas number density within the target clouds (we therefore neglect the temporal, spatial effects within the molecular cloud itself; the whole molecular clouds becomes instantly
a cosmic-ray target). This is a simplification of the model, enough however for the aims herein pursued, which is trying to determine the diffusion environment in the environment, i.e., between the shell and the cloud, outside the latter.}
The CR spectrum is  given by 
\be
   J_p(E,\;r,\;t)= [{c} \beta / {4\pi}] f,
\ee
where $f(E,\;r,\;t)$ is the distribution function of protons at
an instant $t$ and distance $r$ from the source. The distribution
function satisfies the radial-temporal-energy dependent diffusion equation (Ginzburg \&
Syrovatskii 1964):
\be
   ({\partial f}/{\partial t})=({D(E)}/{r^2}) ({\partial}/{\partial
   r}) r^2 ({\partial f}/{\partial r}) + ({\partial}/{\partial
   E}) \, (Pf)+Q,
\ee
where $P=-dE/dt$ is the energy loss rate of the
particles, $Q=Q(E,\;r,\;t)$ is the source function, and $D(E)$
is the diffusion coefficient, for which we assume here that it depends only on the particle's energy. The energy loss rate are due to ionization and nuclear interactions, with the latter dominating over the former for energies larger than 1 GeV. 
{ The nuclear loss rate is $P_{\rm nuc} = E/\tau_{pp}$, with $\tau_{pp}=(n_p\, c \, \kappa \, \sigma_{pp} ) ^{-1}$ being the timescale for the corresponding nuclear loss, $\kappa \sim 0.45$ being the inelasticity of the interaction, and $\sigma_{pp}$ being the cross section (Gaisser 1990). 
Aharonian \& Atoyan (1996) presented a solution for the diffusion equation for an arbitrary energy loss term, diffusion coefficient, and impulsive  injection spectrum $f_{\rm inj}(E)$, such that  $Q(E,r,t) = N_0 f_{\rm inj}(E) \delta{\bar r} \delta(t)$. For the particular case in which $D(E)\propto E^\delta$ and $f_{\rm inj}\propto E^{-\alpha}$,  above $\sim 10$ GeV, where the cross-section to $pp$ interactions is a weak function of $E$,  the general solution is 
\be
  f(E, r,t) \sim ({N_0 E^{-\alpha}}/{\pi^{3/2} R_{\rm dif}^3}) \exp \left[ { - {(\alpha-1)t}/{\tau_{pp} }- ({R}/{R_{\rm dif}})^2} \right],
  \label{sol}
\ee
where $R_{\rm dif} = 2 ( D(E) t [\exp(t \delta / \tau_{pp})-1]/[t \delta / \tau_{pp}])^{1/2}$ stands for the radius of the sphere up to which the particles of energy $E$ have time to propagate after their injection.
 In case of continuous injection of accelerated particles, given by $Q(E, \;
t)=Q_0 E^{-\alpha} {\cal T}(t)$, the previous solution needs to be convolved with the function ${\cal T}(t-t')$ in the time interval $0 \leq t' \leq t$. 
If the source is described by a Heavside function, 
 ${\cal T}(t)=\Theta (t)$ 
Atoyan et al. (1995) have found a general solution for the diffusion equation 
with arbitrary injection spectrum, which with the
listed assumptions and for times $t$ less than the energy loss time, leads to:
\be
f(E,\;r,\;t)=({Q_0 E^{-\alpha}}/{4\pi D(E) r}) (
{2}/{\sqrt{\pi}})\int^{\infty}_{r/R_{\rm diff}} e^{-x^2}
dx.
\ee
We will assume that $\alpha=2.2$ and make use of these solutions in what follows. In Paper I, we gave a detailed description of the multi-frequency knowledge on IC 443, impacting on the determination of the main parameters entering into the model (e.g., the SNR's age, and molecular environment). We refer the reader to that discussion for details.}

\begin{figure*}
\centering
\includegraphics[width=0.45\columnwidth,trim=0 5 0 10]{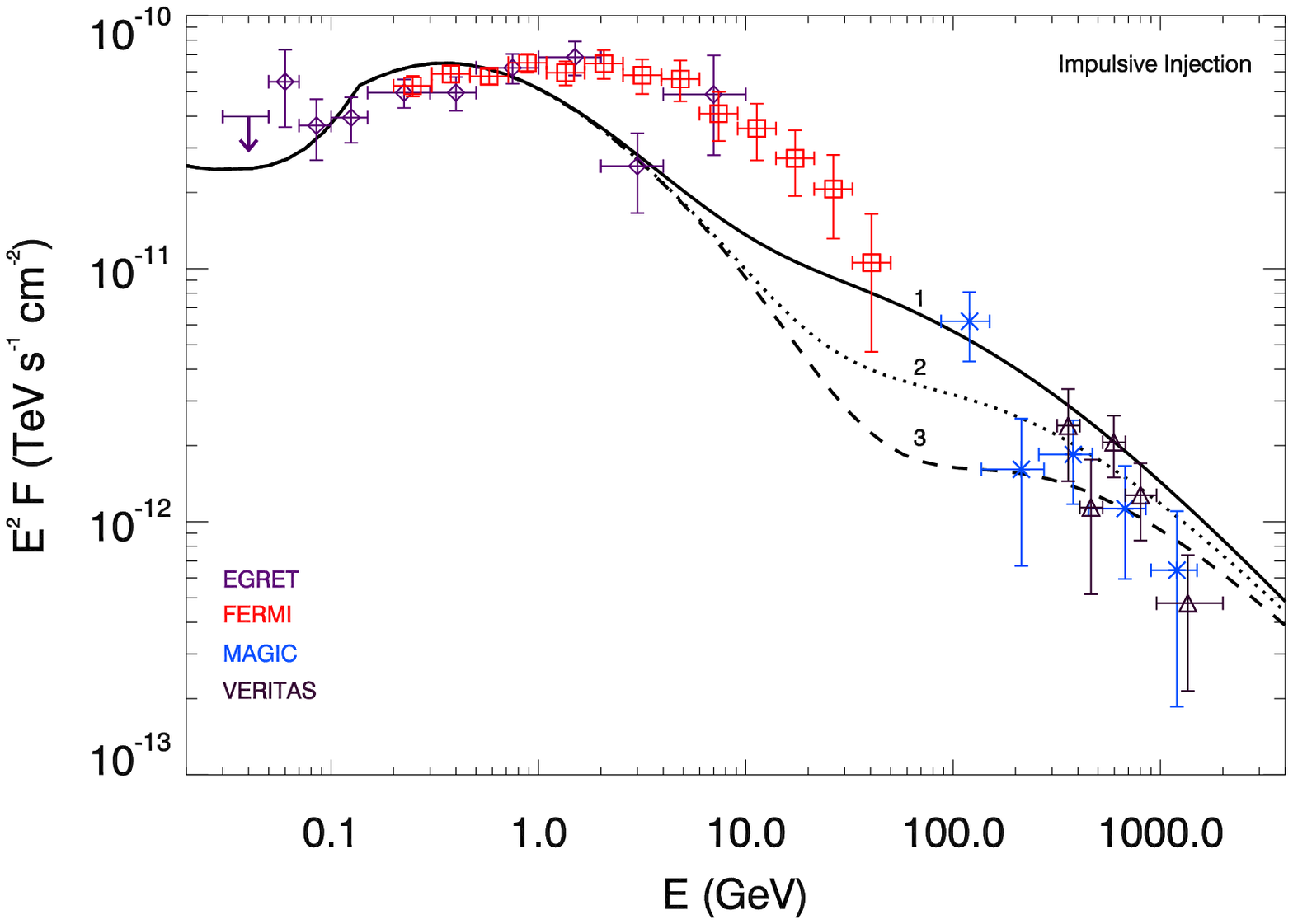}
\includegraphics[width=0.45\columnwidth,trim=0 5 0 10]{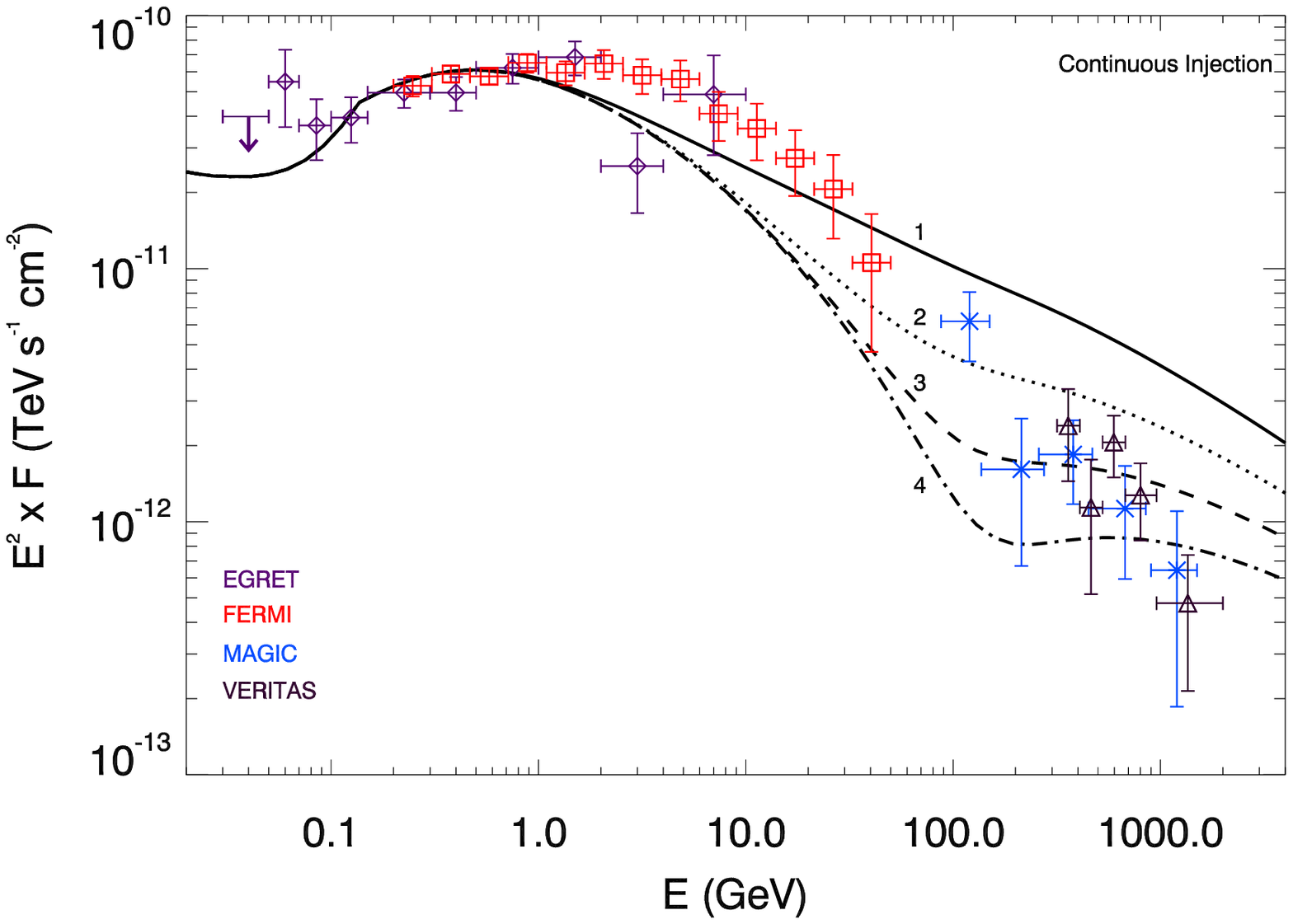}
\caption{Earlier MAGIC and EGRET (stars and diamonds, respectively), and recent {\it   Fermi} and VERITAS (squares and upper trianges, respectively) measurements of the neighborhood of IC 443 as compared with model predictions for an impulsive and a continuous accelerator, as considered in Paper I.
The nominal values of parameters for these models are the following: 
At the MAGIC energy range, the left panel curves show the predictions arising from the $pp$ interactions in a cloud of 8000 M$_\odot$ located at 20 (1), 25 (2), and 30 (3) pc, whereas they correspond to 15 (1), 20 (2), 25 (3), and 30 (4) pc in the right panel. At the EGRET energy range, the curve shows the prediction for a few hundred M$_\odot$ located at 3--4 pc. The right panels show the same results than those in the left, but summing up the different contributions.
}
\label{nominal}
\end{figure*}

\subsection{The model and new data}
\label{dd}

\subsection{Comparison with nominal models in Paper I}

{ We start by directly comparing the predictions made in Paper I with the most recent results obtained by VERITAS and {\it   Fermi}. In the case of VERITAS, given that their measured SED is compatible with the earlier one obtained by MAGIC, we will no see no significant difference in the response of the models. In the case of {\it   Fermi}, the situation is different because {\it   Fermi} results extended the energy  domain of the SED much beyond what was possible for EGRET, and for that region the models explored in Paper I, thus, were unconstrained. }

In the models of Paper I, IC 443 was 
considered both as a continuous accelerator with
a relativistic proton power of  $L_p = 5 \times 10^{37}$ erg s$^{-1}$ 
(the proton luminosity is such that the energy injected into relativistic CRs through the SNR age is $5\times 10^{49}$ erg), and an impulsive injector with the same total power (injection of high energy particles occur in a much shorter time than the SNR age). Cosmic-rays were assumed to propagate with a diffusion coefficient at 10 GeV, e.g., $D_{10}= 10^{26}$ cm$^2$ s$^{-1}$, and $\delta=0.5$ in a medium of  typical density, for which the timescale for nuclear loss $\tau_{pp}$ is orders of magnitude larger than the age of the accelerator.
The nominal models explored also took assumptions regarding the location (different distances between the SNR shock and the interacting clouds were assumed) and molecular mass affected by the cosmic-rays.  These assumptions were based on the observations of molecular lines towards IC 443 made by, e.g., 
Cornett et al. (1977), De Noyer (1981), Dickman et al. (1992), Seta et al. (1998), and Torres et al. (2003) which conform the overall picture: 
a total mass of $\sim 1.1 \times 10^4$ M$_\odot$ mostly at the foreground of the remnant,  since it is found to be absorbing optical and X-ray radiation, with smaller cloud(s) totalizing the remaining mass located closer to the SNR. Important details are however uncertain, for instance,  whether there is one or several foreground clouds, the distance between the foreground cloud(s) and the SNR shell, the number and specific location(s) of the foreground cloud(s), and their mass distribution if more than one cloud is there. It was the hoped that {\it   Fermi} data would elucidate some of these parameters, { regarding not only the molecular environment but also the diffusion properties of the medium, by a posteriori comparison with data. } 

Figure \ref{nominal} shows the result of the nominal model predictions (theoretical curves are exactly as in Paper I, except for the fact we computed and added a contribution of bremsstrahlung radiation { (we consider this contribution for primary particles with a proton to electron ratio of 150, following the standard formulae quoted for instance in Torres 2004)} which is only visible at the smallest energies in the plot, compared with the newest data. { We note that electron bremsstrahlung can hardly explain the whole of the observed IC 443 gamma-ray emission, a conclusion also reached by Abdo et al. (2009), and with EGRET data, also by Butt et al. (2003). Since the cross section of bremsstrahlung and pion production are similar at the {\it Fermi} range,  their emission ratio is similar to the electron-to-proton ratio. }
The curves in Figure \ref{nominal} are based on assuming 8000 M$_\odot$ at the different distances marked in the plot and a few hundred M$_\odot$ located closer to the SNR (as an example, $\sim$700 M$_\odot$ for the case of an impulsive, and $\sim$300 M$_\odot$ for a continuous case, located at 3--4 pc). One can immediately see that what earlier was, particularly in the case of the impulsive accelerator, a good agreement between theory and the observations performed by EGRET and MAGIC (and also VERITAS) { only},  is now in disagreement with {\it   Fermi} data. The spectrum is harder than what was suggested by EGRET, presenting an almost flat SED up to 10 GeV, with a roll-over in the spectrum between 10 and 100 GeV. Models in Paper I are unable to reproduce the details of these trends: { In fact, } the case of continuous acceleration was already not favored in Paper I due to both, the middle age of the remnant and the behavior at the highest energies, which were producing a SED much harder than observed { and it is now ruled out}. We will not consider this case any further. In the case of impulsive acceleration, it is at the earlier unexplored region of energies, between 10 and 100 GeV, where we find significant deviations between theory and data, { and there is no model among the ones explored above } which can accommodate at { the same} time a SED that is both, sufficiently steep at VHEs to concur with MAGIC/VERITAS observations and sufficiently flat one decade earlier in energy to concur with {\it   Fermi} data.

\subsection{Using {\it   Fermi} data to constrain model parameters}

\begin{figure*}
\centering
\includegraphics[width=0.45\columnwidth,trim=0 5 0 10]{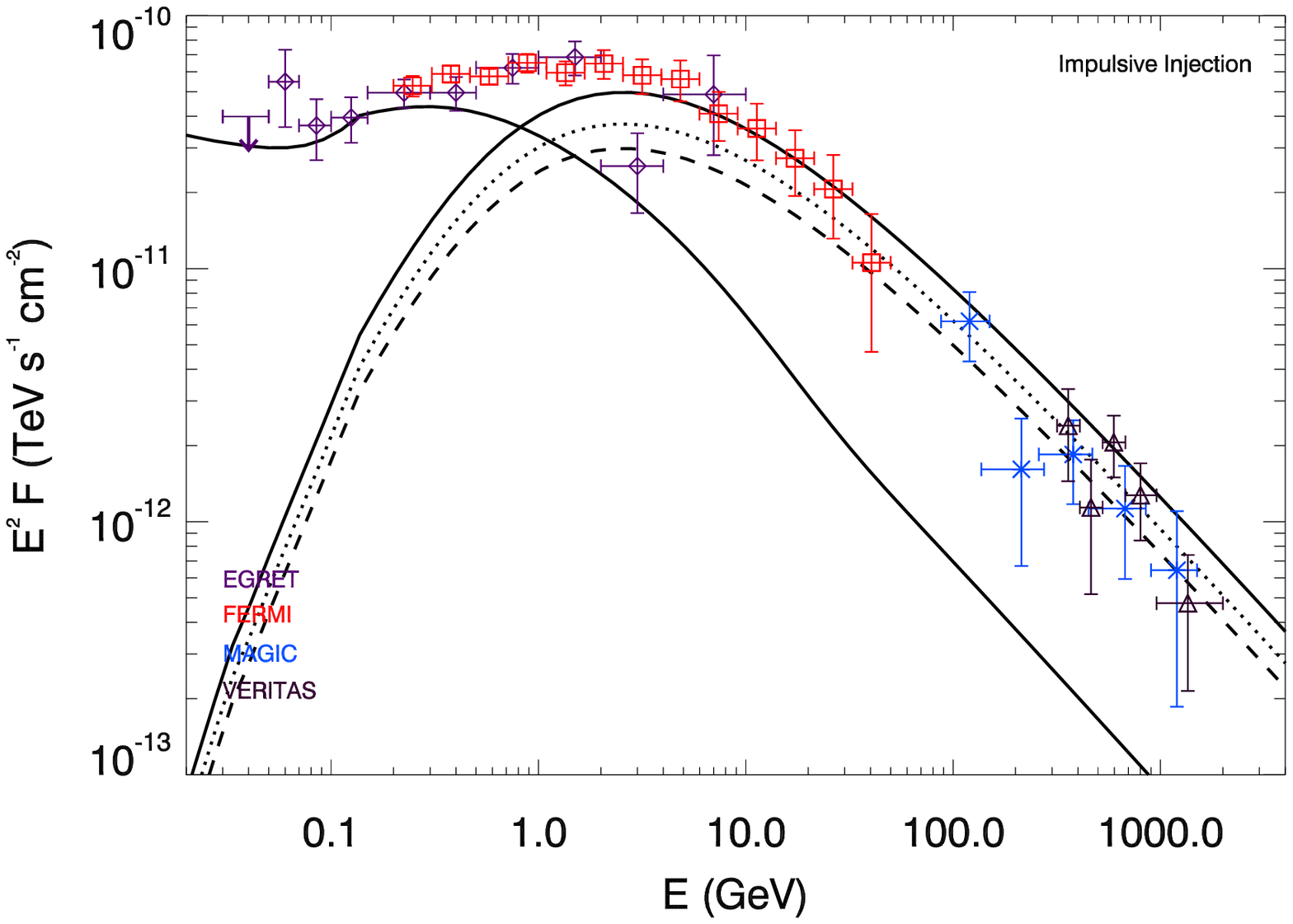}
\includegraphics[width=0.45\columnwidth,trim=0 5 0 10]{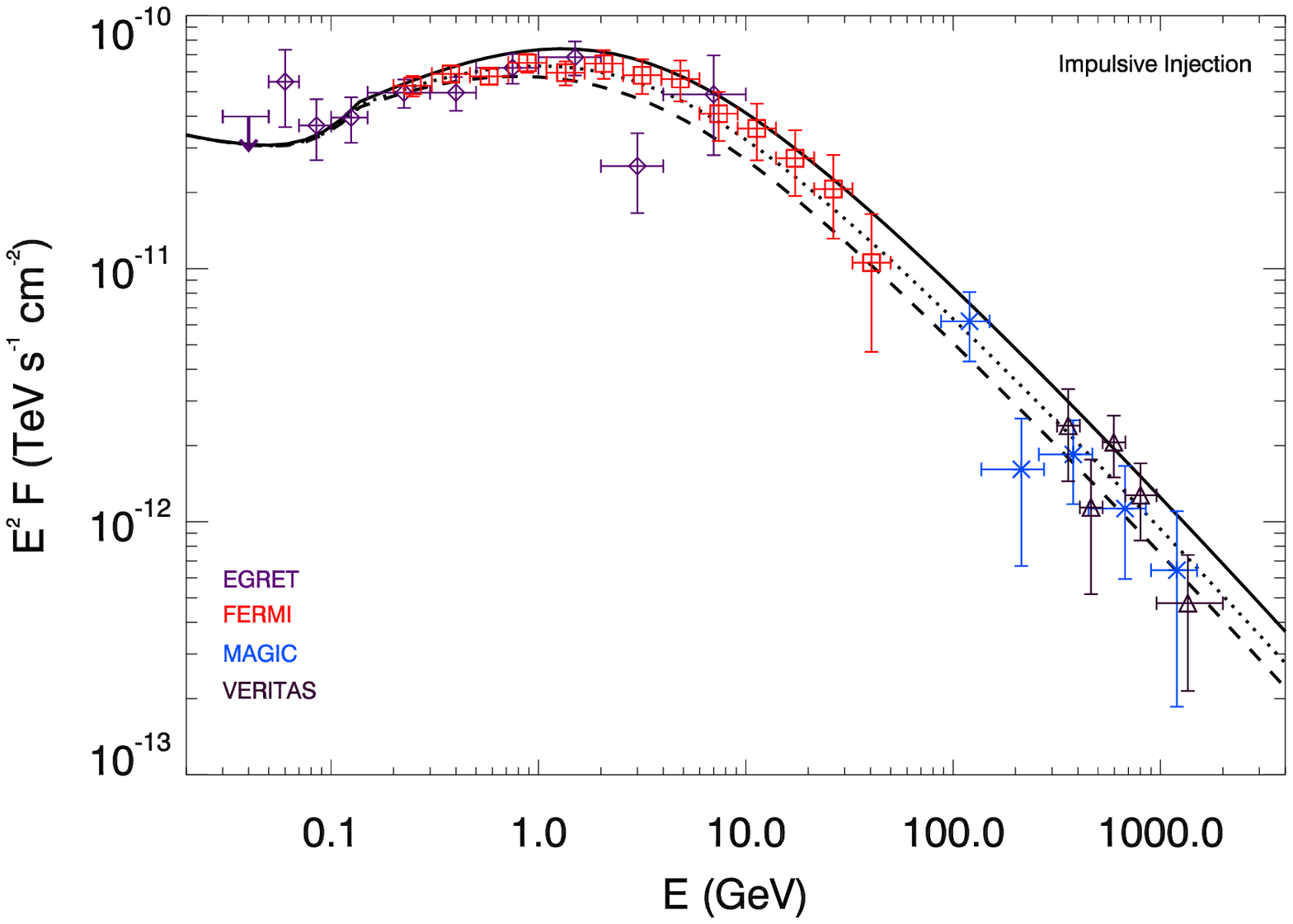}
\caption{As in Figure 1. Summed results (right) are produced by two main components (shown in the left panel) coming from a giant cloud in front of the SNR, which is at least partially overtaken by the diffusing cosmic-rays ($\sim$5300 M$_\odot$ at 10 pc) and a closer-to-the-shell cloud, similar to the previous examples (in this case, at 4 pc, with 350  M$_\odot$). The dotted (dashed) line at the VHE range corresponds to different normalizations (equivalently, to interacting masses of $\sim$4000 and $\sim$3200 M$_\odot$ at the same distance). The diffusion coefficient is as before, $D_{10}= 10^{26}$ cm$^2$ s$^{-1}$.}
\label{new-model}
\end{figure*}

{ What at first sight could seem as a difficult-to-solve failure of the scenario, we find that it is actually only the failure of some numerical values of  parameters.  In particular, }
differences in the location and masses of the overtaken clouds can move the peaks of their corresponding contributions (see Aharonian \& Atoyan 1996, { Gabici et al. 2007, Rodriguez-Marrero et al. 2009  
for detailed analysis of the dependences).} 
{ Certainly, kinematic distance estimations are not accurate enough to obtain exact separation of the cloud(s) from the SNR shell. Thus, {\it   Fermi} observations are holding the key to make some precisions on the assumptions made in this sense, given that the unknowns can affect the final results on the predicted spectra. }
Using {\it   Fermi} results we find that a closer (e.g., at 10 pc) less massive giant cloud ($\sim$5300 M$_\odot$) being overtaken by cosmic-rays diffusing away from IC 443 and an smaller amount of molecular material in cloud(s) closer to the SNR shell (e.g., at 4 pc, with 350  M$_\odot$) produce an excellent match to the whole range of observations, see Figure \ref{new-model}.
{ A smaller --than the total quoted: $1.1 \times 10^4$ M$_\odot$-- amount of mass in the foreground giant molecular cloud(s) being overtaken by diffusing cosmic-rays from IC 443 is perfectly possible, given the various uncertainties in the absolute position of the cloud, its real number, and the velocity model used; and essentially, due to the fact that the total amount of mass correspond to a larger projected sky area. Whereas this implies no substantial change to the model, it allows a match with data at all high-energy frequencies, see Figure \ref{new-model}.
As in Figure 1, this Figure shows the results produced by two main components, one coming from a giant cloud in front of the SNR, which is at least partially overtaken by the diffusing cosmic-rays (e.g., $\sim$5300 M$_\odot$ at 10 pc) and a closer-to-the-shell cloud, of overall magnitudes similar to the previous examples (in this case, at 4 pc, with 350  M$_\odot$). The dotted (dashed) line at the VHE range corresponds to different normalizations (equivalently, to interacting masses of $\sim$4000 and $\sim$3200 M$_\odot$ at the same distance). The diffusion coefficient is as before, $D_{10}= 10^{26}$ cm$^2$ s$^{-1}$.}

\subsection{Cosmic-ray distributions and their effects}

\begin{figure}
\centering
\includegraphics[width=0.45\columnwidth,trim=0 5 0 10]{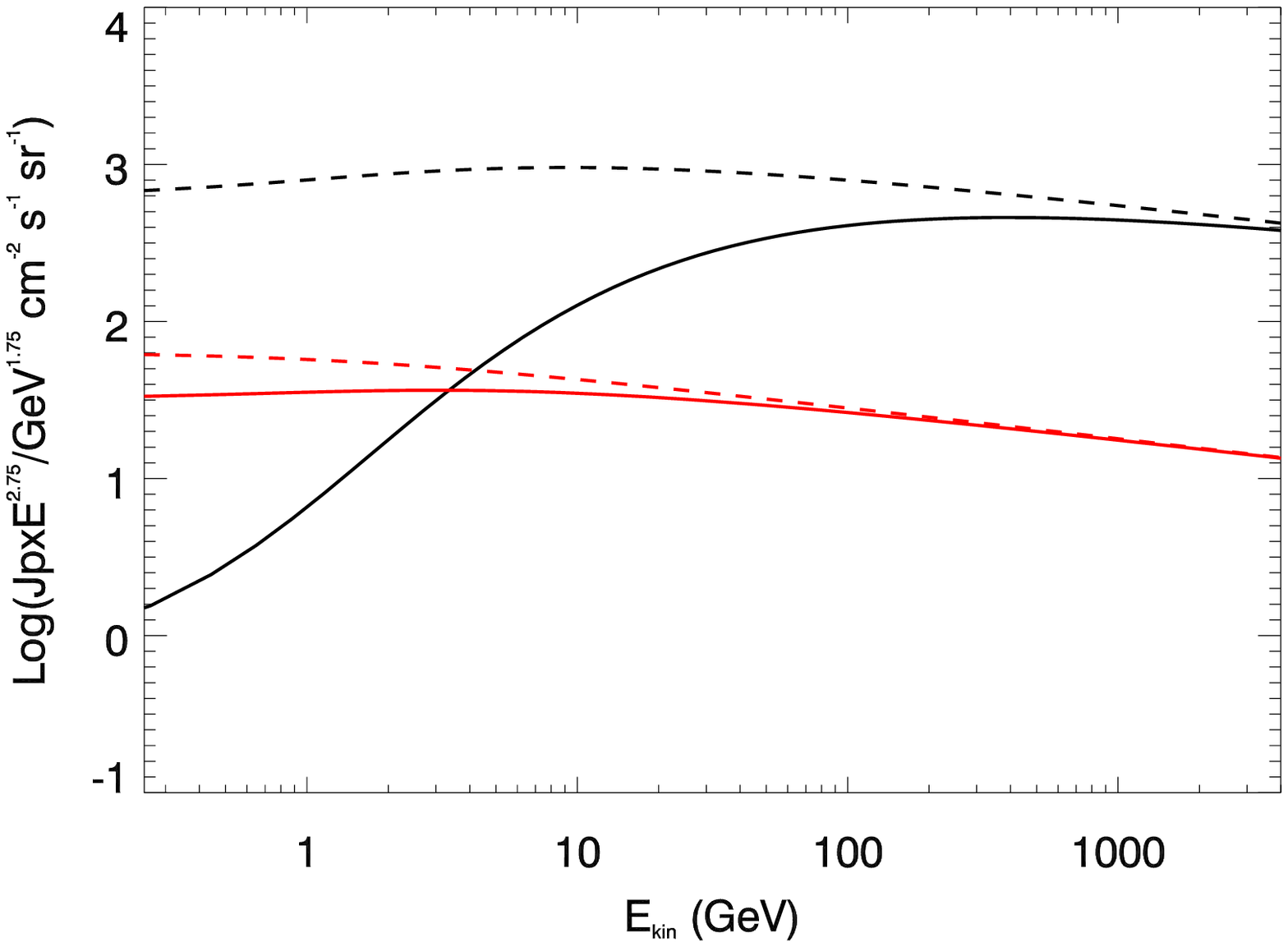}
\includegraphics[width=0.45\columnwidth,trim=0 5 0 10]{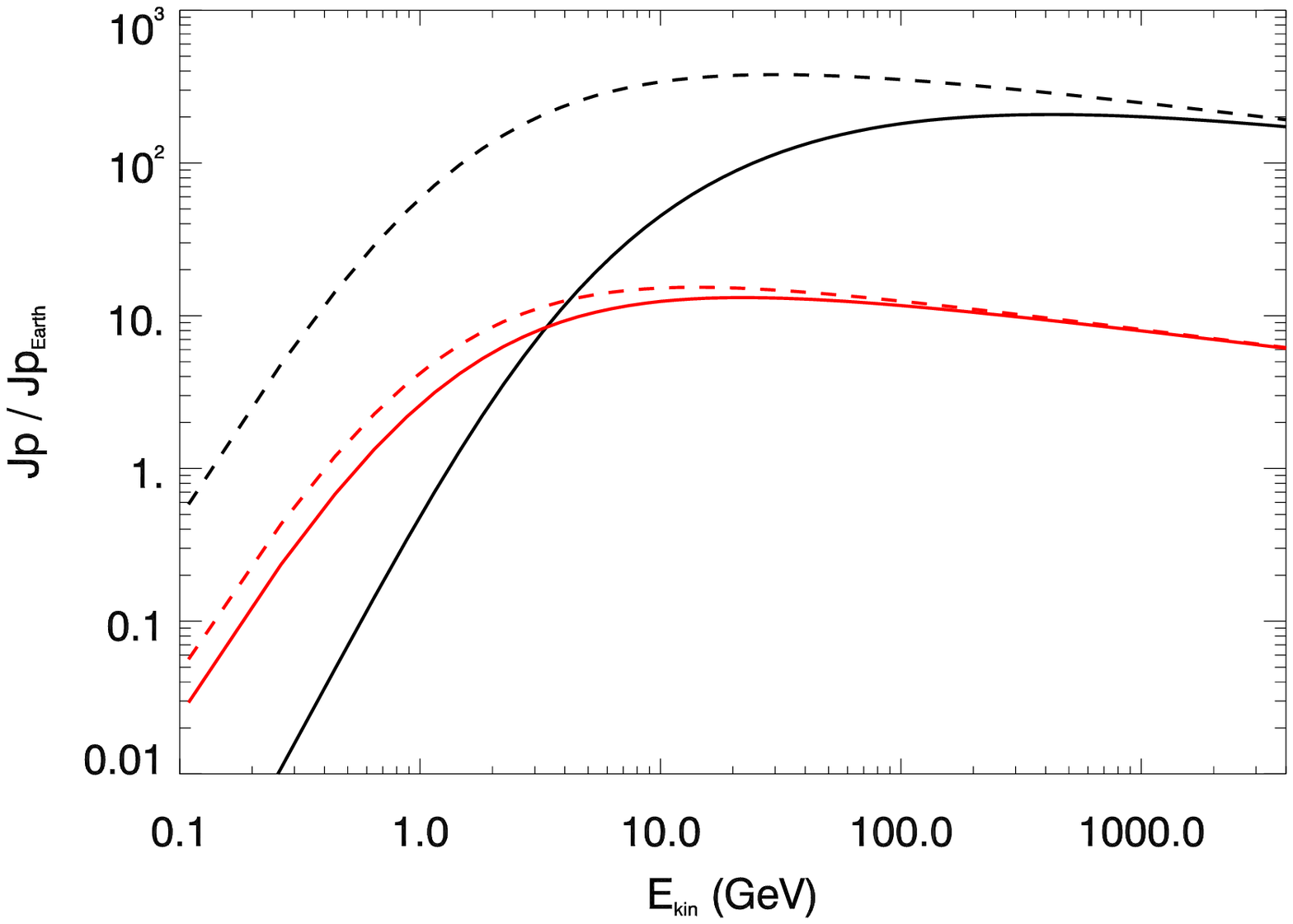}
\caption{Cosmic-ray spectrum generated by the impulsive IC 443 at the two different cloud distances considered in one matching model of Figure \ref{new-model}: 10 (solid) and 4 pc (dashed), at the age of the SNR, as a function of energy. Different colors show results for different diffusion coefficient (black, $D_{10}= 10^{26}$ cm$^2$ s$^{-1}$; and red, $D_{10}= 10^{27}$ cm$^2$ s$^{-1}$). 
{ The right panel shows the ratio between the cosmic-ray spectra of the top panel, and the cosmic-ray spectrum near Earth, as a function of energy.}
}
\label{new-model-CR}
\end{figure}

\begin{figure}
\centering
\includegraphics[width=0.45\columnwidth,trim=0 5 0 10]{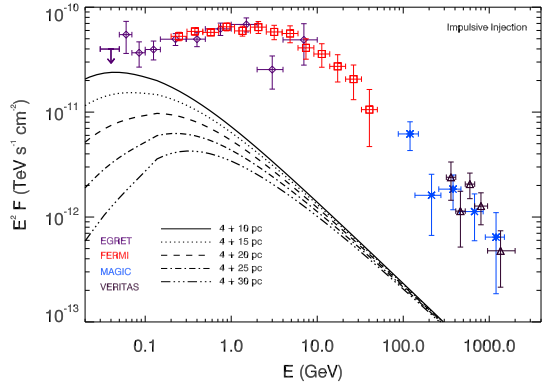}
\caption{Example of model output with diffusion coefficient scale equal to $10^{27}$ cm$^2$ s$^{-1}$. The different curves represent results for the location of giant molecular cloud at 10, 15, 20, 25, 30 pc from the SNR shell, whereas the close-to-the-SNR cloud is at 4 pc. 
Neither in this nor in any other of the models studied varying the parameters with such diffusion coefficient scale, the VHE source spectrum can be reproduced, nor the resulting SED in {\it   Fermi} range is hard enough to match the data. }
\label{D27}
\end{figure}

In Figure \ref{new-model-CR} we show the distribution of cosmic-rays generated by the impulsive IC 443 at the two different distances considered for the molecular mass distribution in one matching model { (solid black line)} of Figure \ref{new-model}. We also plot their ratio with respect to the Earth cosmic-ray distribution. It can be seen that the cosmic-ray energy density is greatly enhanced --along the energy range of interest--- as compared with that in our vicinity, described with an spectrum of the form $J_\odot (E) \sim 2.2
E_{\rm GeV}^{-2.75}$ cm$^{-2}$ GeV$^{-1}$ s$^{-1}$ sr$^{-1}$ { (e.g. Dermer 1986)}. 
It can also be seen that significant deviations of the cosmic-ray density are obtained in case the diffusion is slower (i.e., $D_{10}$ is larger). At a fixed SNR age of 30 kyrs, 
increasing $D_{10}$  produces the $\gamma$-ray emission prediction to displace 
to smaller energies,
{ typically, until $D_{10} > D_{transition}$, where peaks generated by clouds at large separation (e.g, 100 pc) displace up and peaks generated by clouds at smaller separation (e.g., 10 pc) displace down in the SED 
(e.g., Rodriguez-Marrero et al. 2009 and references therein). }
This fact implies that for the range of distances to the giant and close-to-the-SNR molecular clouds being considered (10--30 pc, and 2--6 pc respectively) there is no solution { with large $D_{10}$} able to fit the whole range of data. This was already hinted at in Paper I, where just using MAGIC data
we found that it was possible to put an strong constraint over the diffusion timescale:  $D_{10}$ should be of the order of 10$^{26}$ cm$^2$ s$^{-1}$ since 
if the separation between the giant cloud and the SNR is $>$10 pc, an slower diffusion would not allow sufficient high energy particles to reach the target material and it would be impossible to reproduce the VHE data. On the other hand, given that there is a displacement between the centroid positions of EGRET/{\it   Fermi} and VHE sources and that molecular material is absorbing lower frequency emission from the remnant, the separation between the foreground cloud(s) and SNR shell can not be much smaller than 10 pc. The current {\it   Fermi} data emphasizes this conclusion.
{ Figure \ref{D27} shows an example of a full range of models we constructed with $D_{10}=10^{27}$ cm$^2$ s$^{-1}$, and its disagreement with data. Note that even if rescaling the curves assuming e.g., a much higher molecular mass (which would in itself be in conflict with multi-frequency observations), it is not possible to obtain a good fit across the whole range of observations.}

\begin{figure}
\centering
\includegraphics[width=0.6\columnwidth,trim=0 5 0 10]{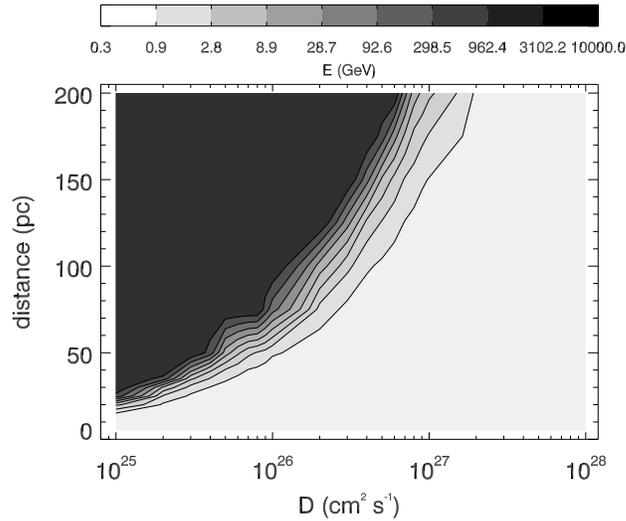}
\caption{Contour plot depicting the position of the peak of the SED generated by a 30000 years old injection interacting with clouds at 
different distances, for a range of diffusion coefficient scale, $D_{10}$}
\label{contour}
\end{figure}

Figure \ref{contour} shows, as contour plots, the energy at which the maximum of the SED is found for the cases of impulsive  acceleration of cosmic-rays. The age corresponds to the  SNR like IC 443, and the cosmic-rays are interacting with clouds at 
different distances, for a range of diffusion coefficient scale, $D_{10}$. This plot is useful to notice which is the solution we are finding and its { degree of uncertainty/degeneracy}: in order to fit the combined MAGIC/VERITAS and {\it   Fermi} data we would need a giant cloud producing a peak at about the {\it   Fermi} spectral turnover (what means, looking at the plot, { either} a very large separation between the cloud and the SNR shell for a high $D_{10}$, (what we discarded because it is not possible to fit the VHE MAGIC/VERITAS data in this configuration), { or a smaller distance with a faster diffusion, i.e. a lower $D_{10}$, which is the solution we can still promote.}

\subsection{More on degeneracies and uncertainties in parameter estimation}

\begin{figure*}
\centering
\includegraphics[width=0.33\columnwidth,trim=0 5 0 10]{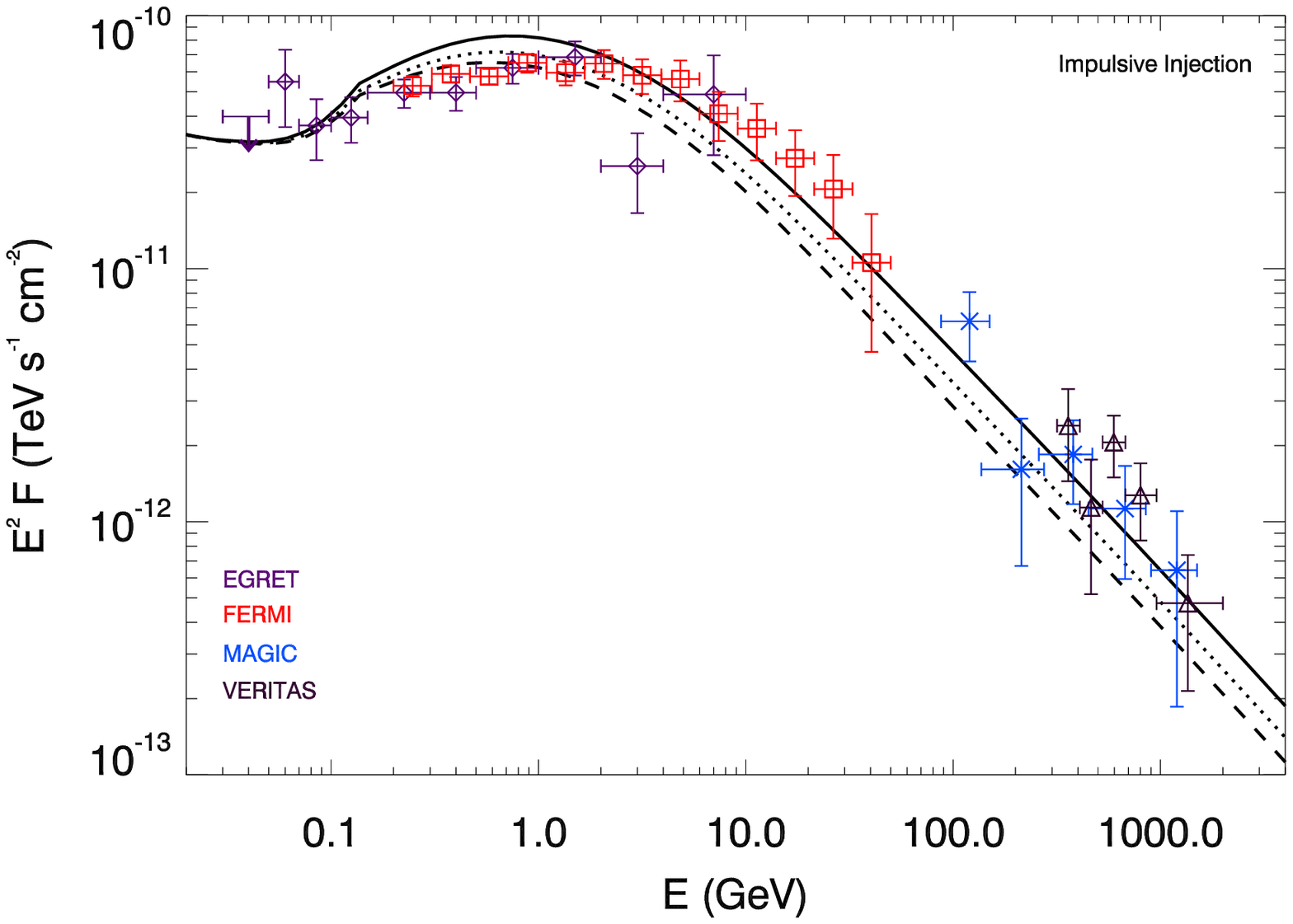}
\includegraphics[width=0.33\columnwidth,trim=0 5 0 10]{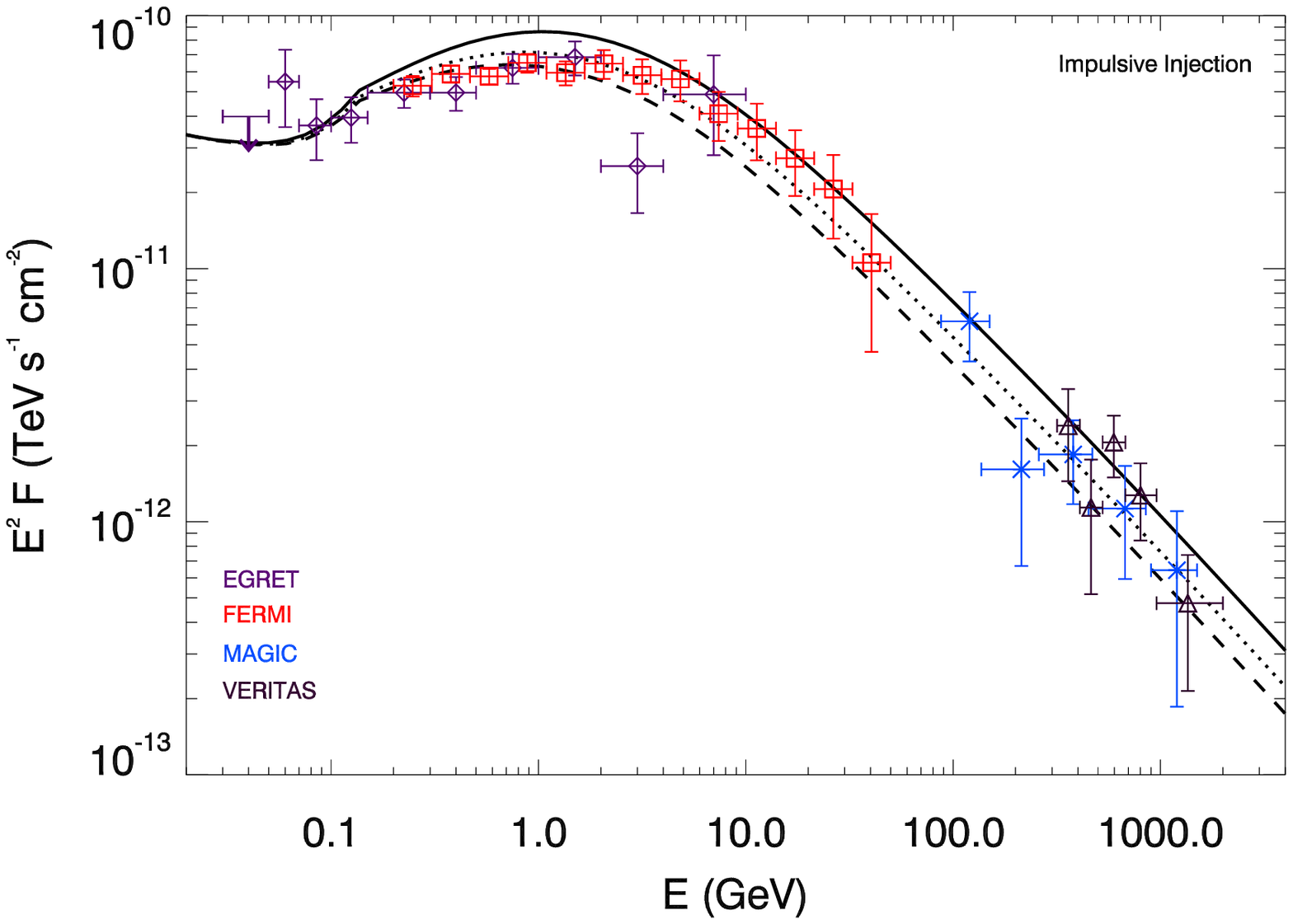}
\includegraphics[width=0.33\columnwidth,trim=0 5 0 10]{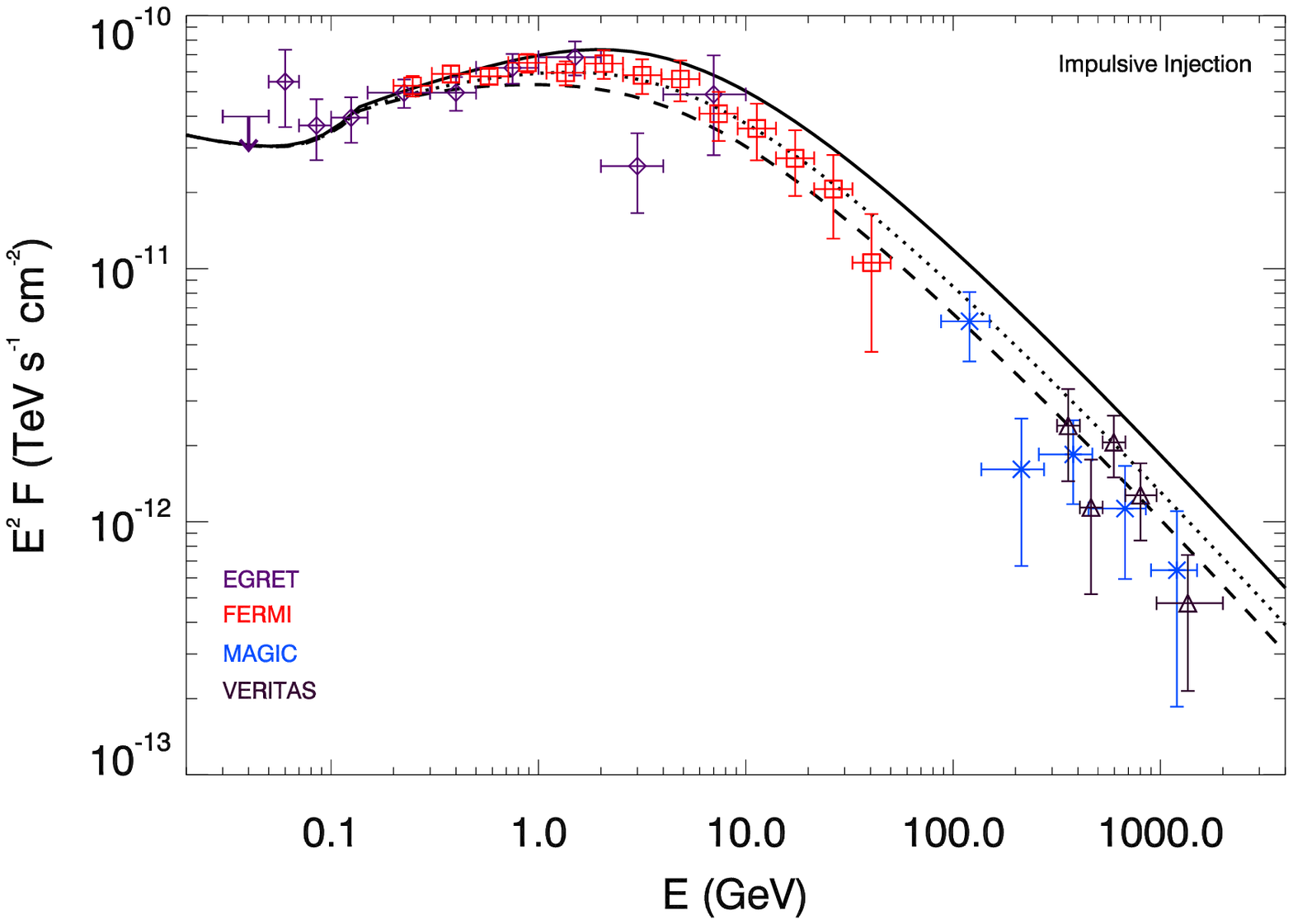}
\includegraphics[width=0.33\columnwidth,trim=0 5 0 10]{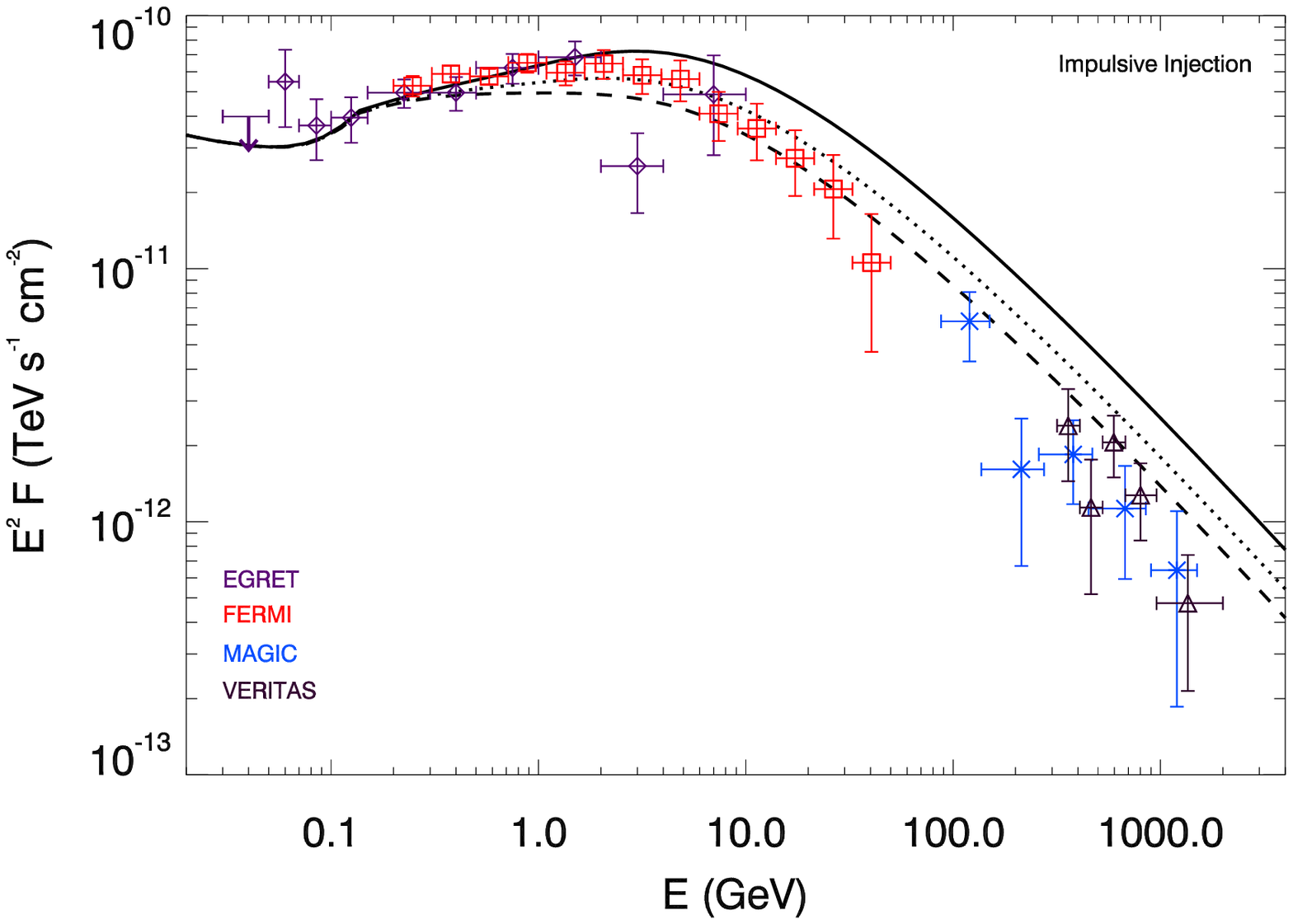}
\includegraphics[width=0.33\columnwidth,trim=0 5 0 10]{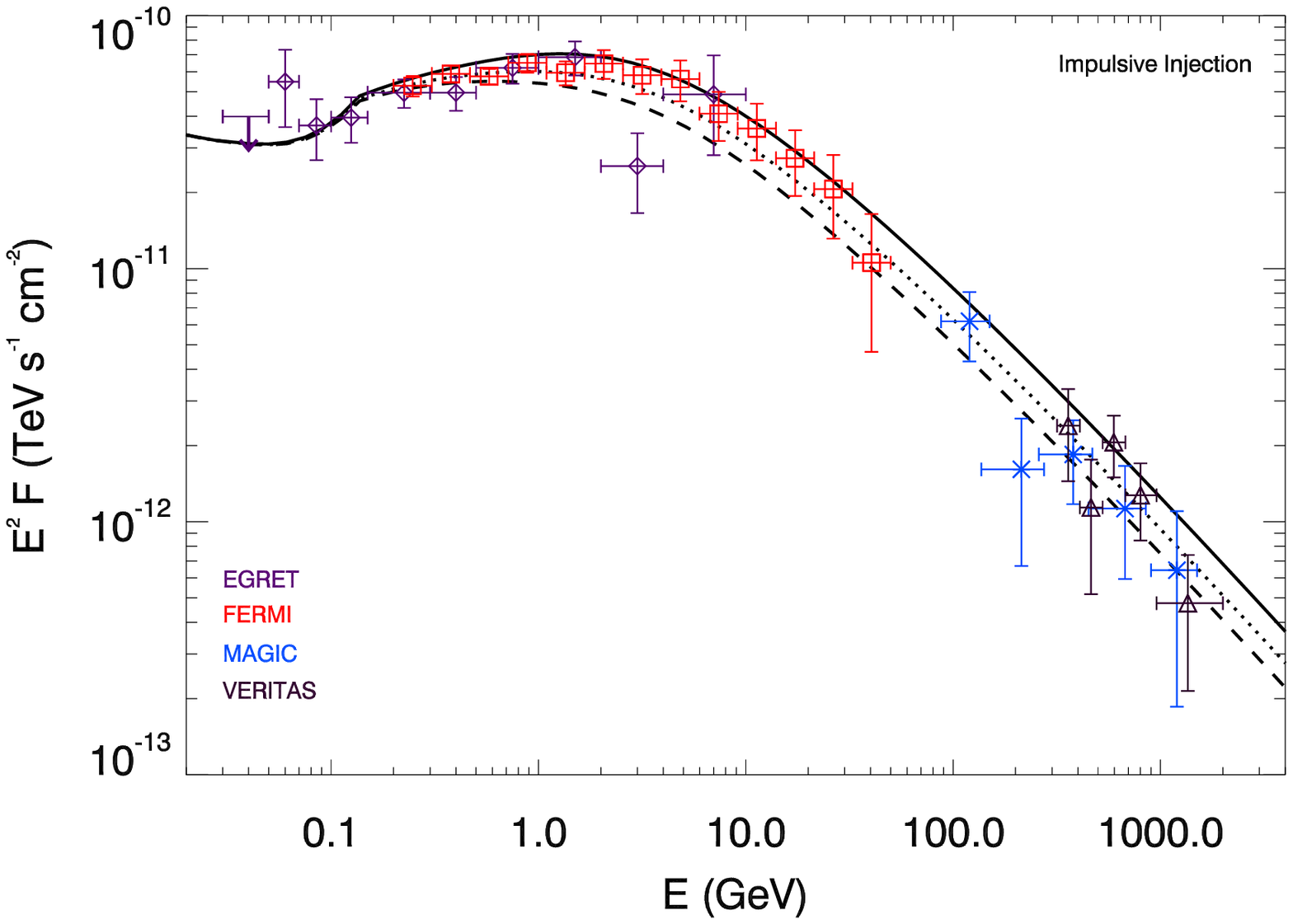}
\includegraphics[width=0.33\columnwidth,trim=0 5 0 10]{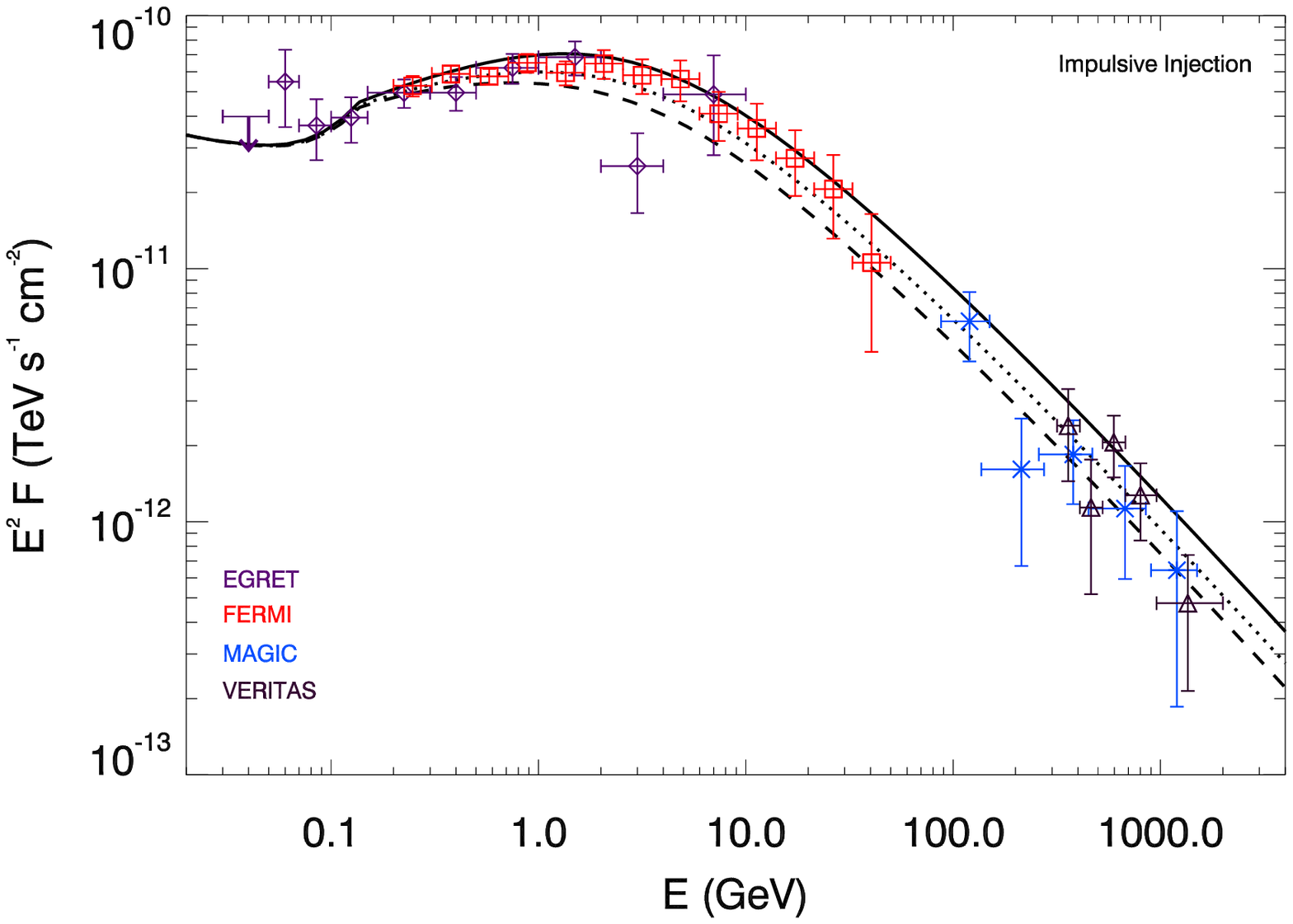}
\includegraphics[width=0.33\columnwidth,trim=0 5 0 10]{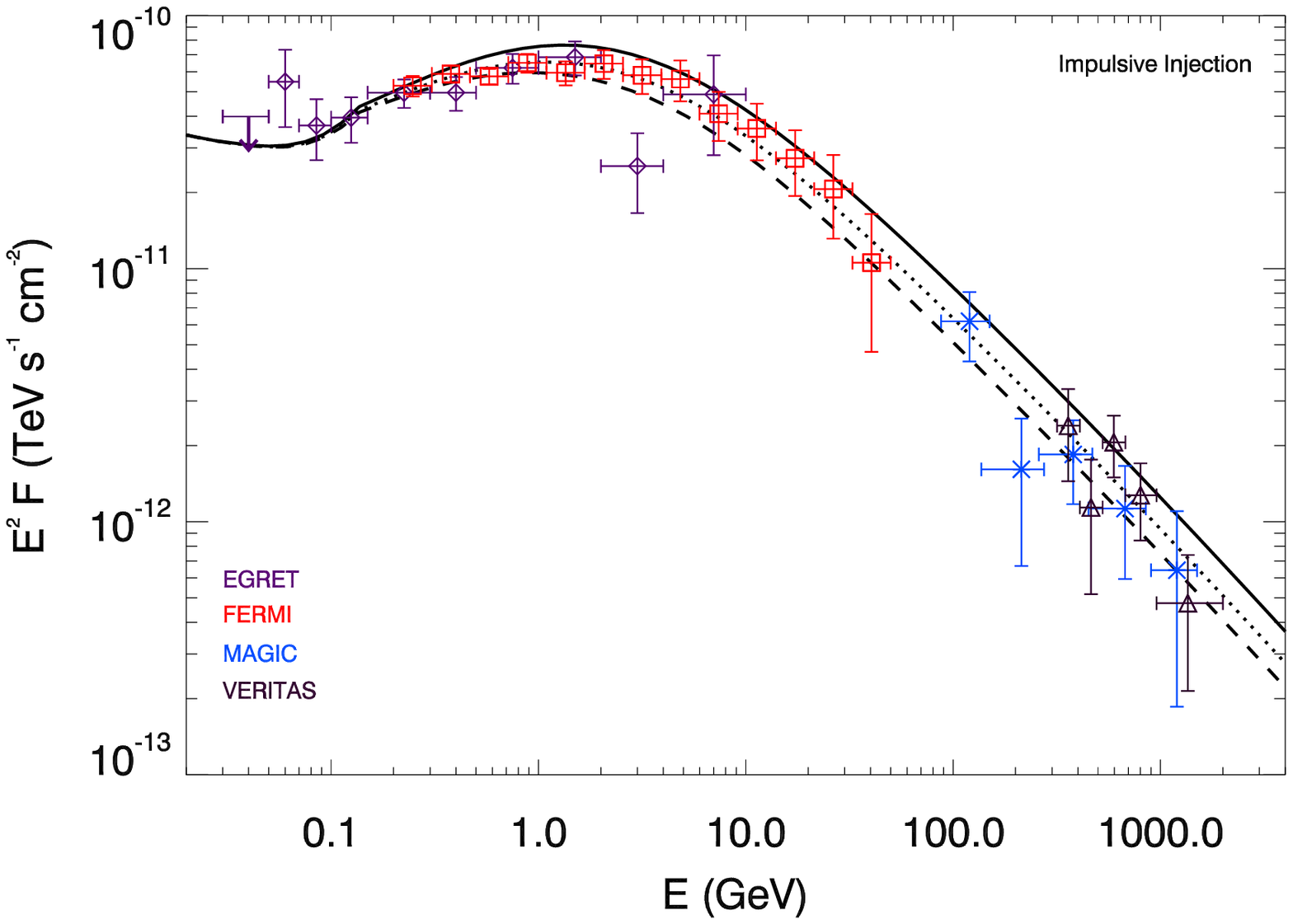}
\includegraphics[width=0.33\columnwidth,trim=0 5 0 10]{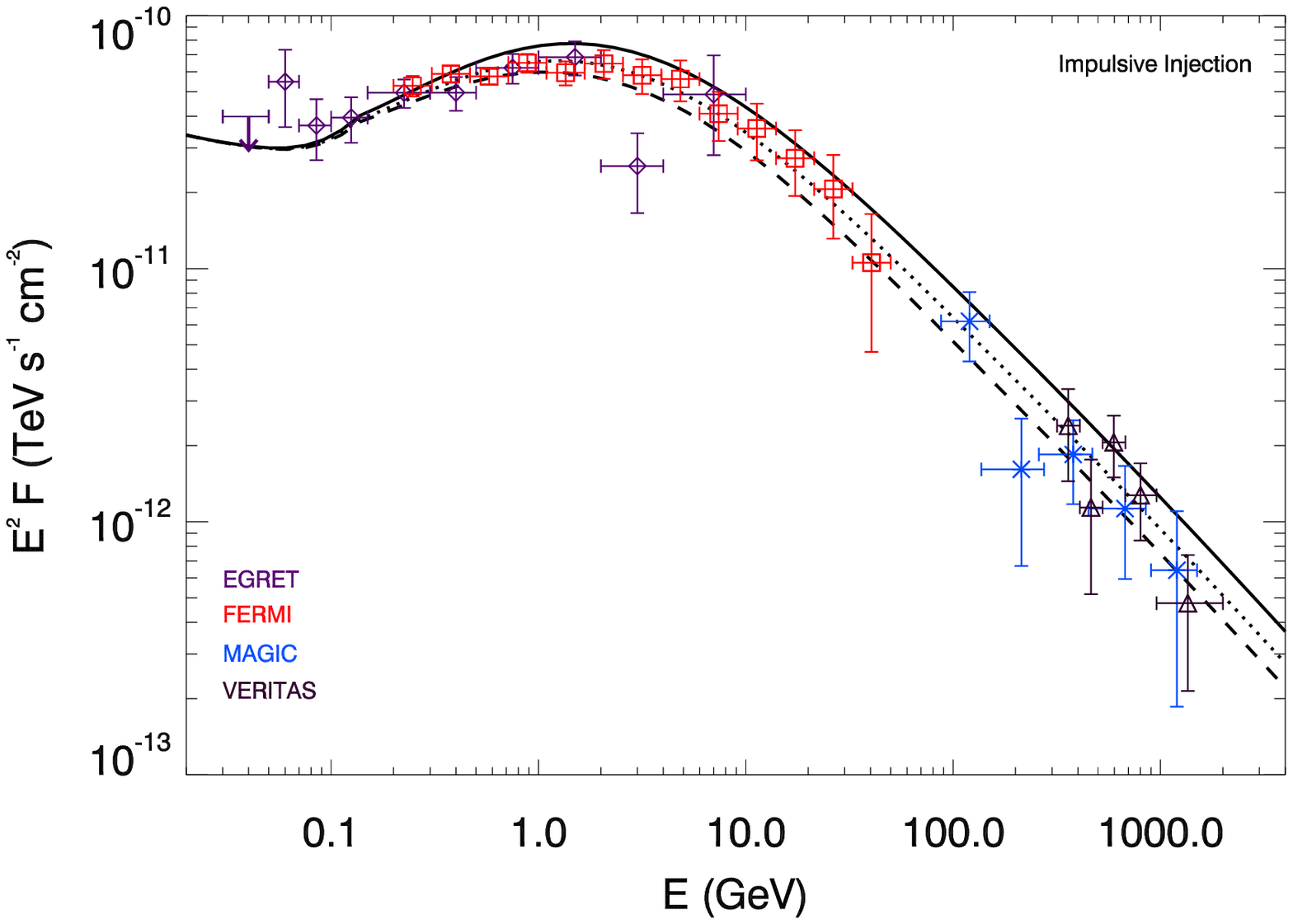}
\includegraphics[width=0.33\columnwidth,trim=0 5 0 10]{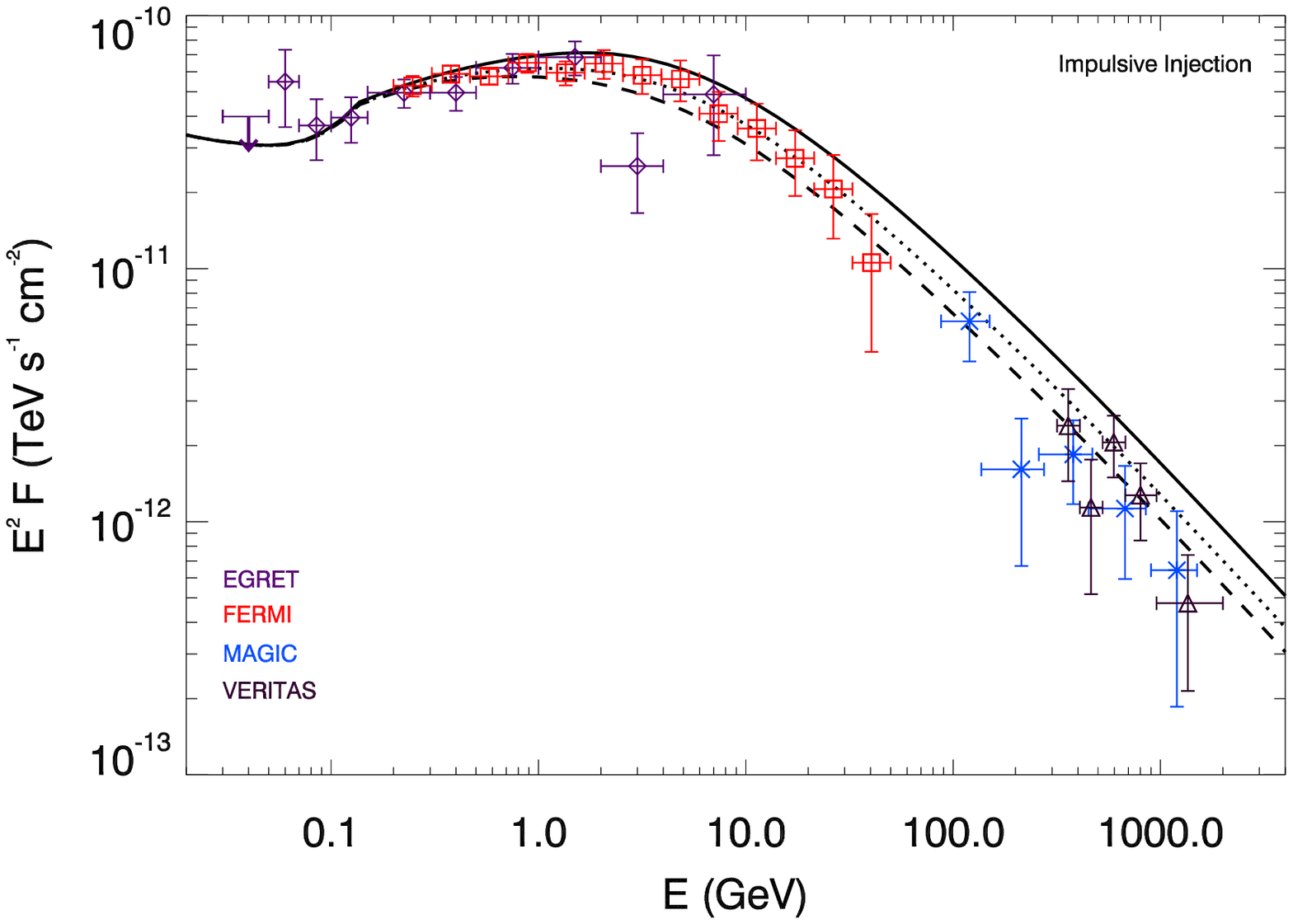}
\includegraphics[width=0.33\columnwidth,trim=0 5 0 10]{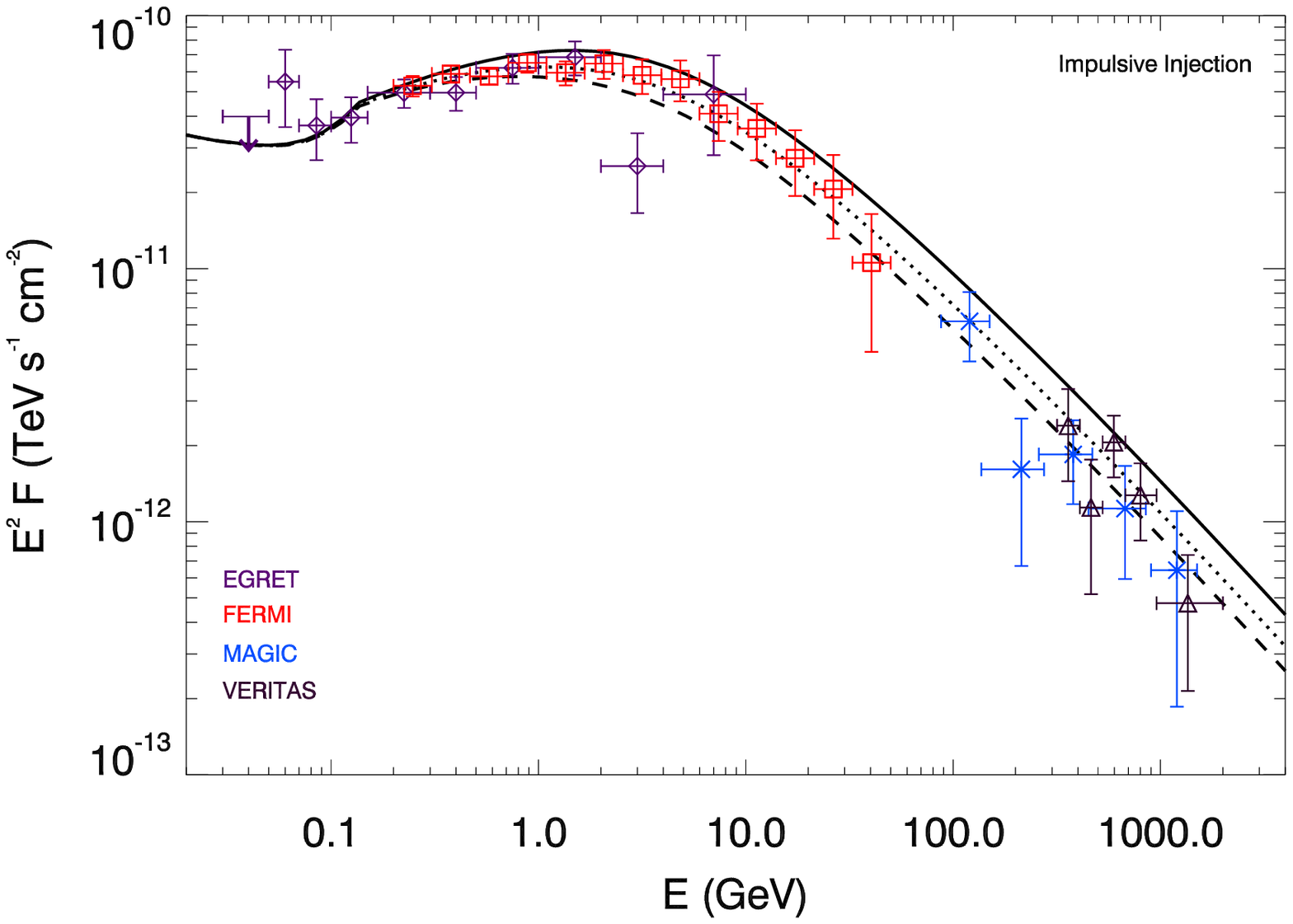}
\includegraphics[width=0.33\columnwidth,trim=0 5 0 10]{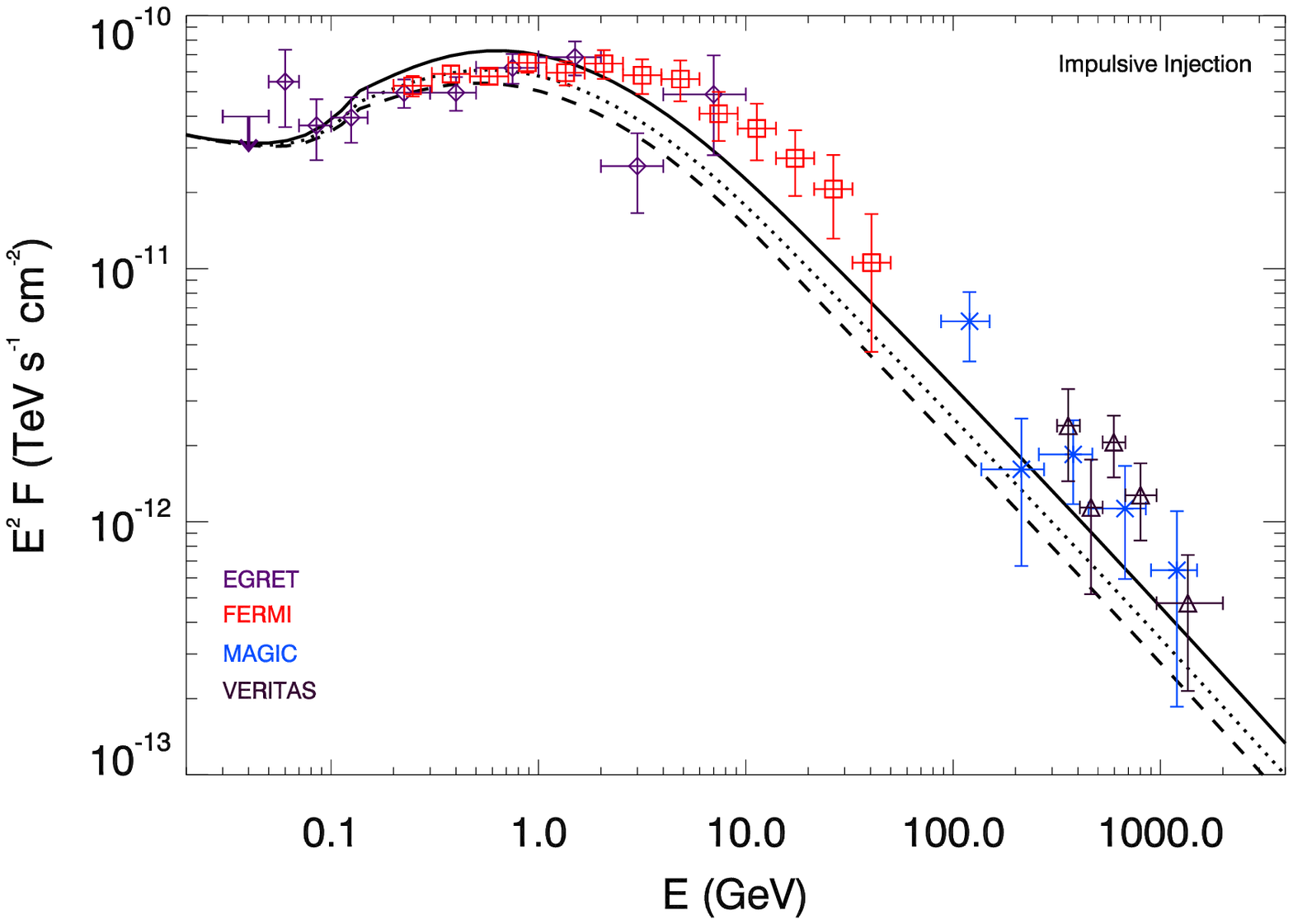}
\includegraphics[width=0.33\columnwidth,trim=0 5 0 10]{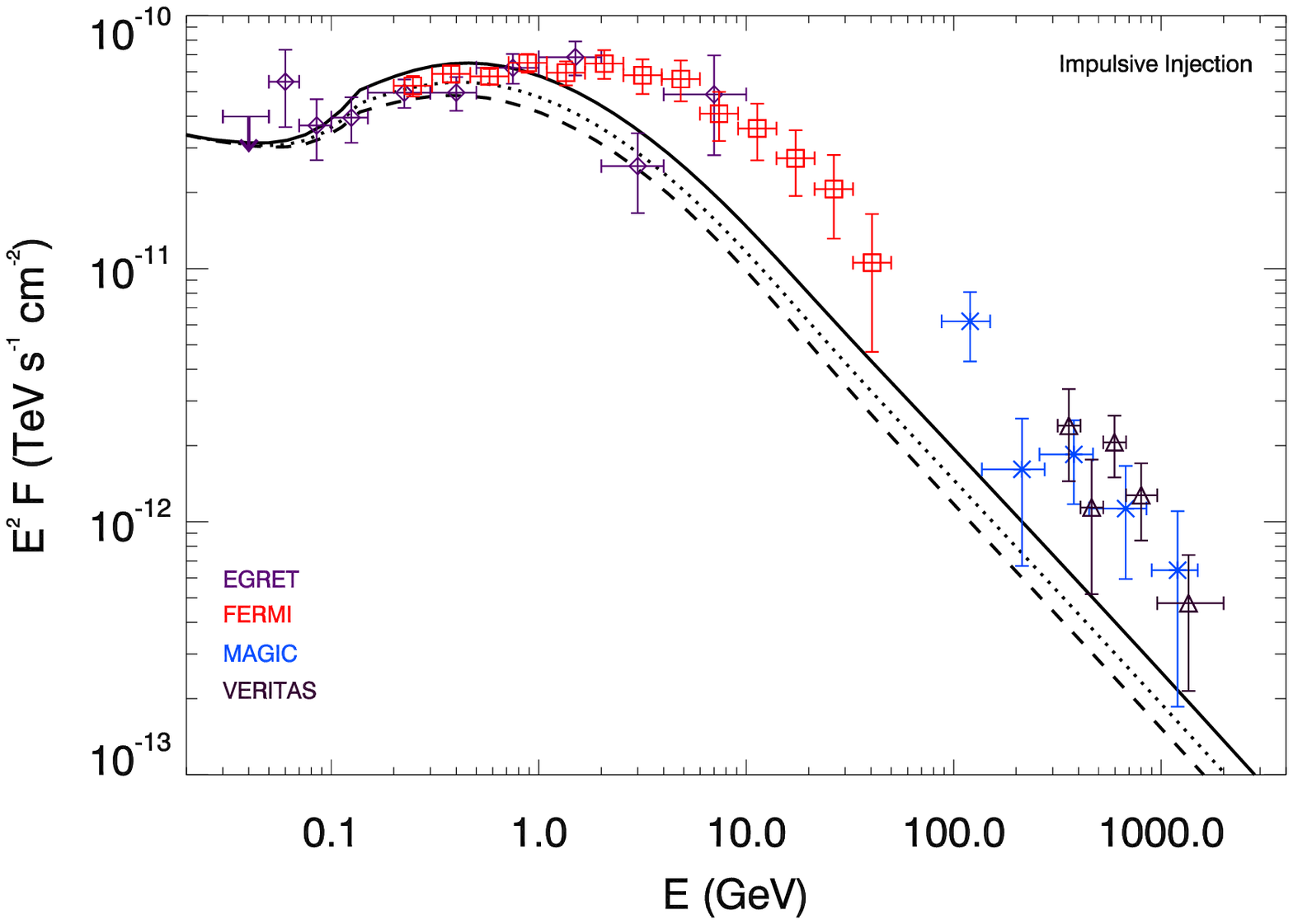}
\caption{Examples of solutions around the main values discussed, exploring the degeneracies (or uncertainties) in determining the numerical values of model parameters matching the observational data. The order of the panels in this plot, top left to bottom right,  corresponds with the parameters
described in Table 1.
}
\label{deg}
\end{figure*}

Figure \ref{deg} explores the range of parameters around the solutions matching the observational data; giving a feeling of the degeneracies (or uncertainties) within which this model provides a reasonable agreement with observations. The values of masses and diffusion coefficients used in Figure \ref{deg} to obtain good data-matching given the distances to each of the clouds are given in Table 1.
Fits could be considered good for $D_{GMC}$ 
between 9 and 11 pc. For an average distance of 10 pc, the mass in the close-to-the-SNR cloud (or clouds) decreases the farthest the latter is. For these cases, good solutions with $D_{GMC}$ 
between 9 and 11 pc can always be found adjusting other parameters. 
Our average model explored in Figure \ref{new-model} corresponds to 
$D_{GMC}=10$ pc,  $d_{snr}$=4 pc, with the three curves constructed with $\sim$5300, $\sim$4000, $\sim$3200 M$_\odot$, and $M_{snr}=350$ M$_\odot$. The results in Figure \ref{deg} and Table 1 show that the smaller the diffusion coefficient, the fit at VHEs worsens, overpredicting the data. Correcting this via a mass adjustment, would in turn make for a poor fit at lower energies; what in practice imposes a lower limit to $D_{10}$. On the other hand, when $D_{10}
$ increases the VHE spectra is quickly underpredicted, and again,  correcting this via a mass adjustment would in turn make for a poor fit at lower energies. In summary, in order for this model to match the multi-frequency observational data, the range of variation in the parameters gets constrained as $9  \lesssim D_{GMC} \lesssim 11$ pc; 3 $ \lesssim d_{snr} \lesssim 6$ pc,
and $D_{10} \sim 10^{26}$ cm$^2$ s$^{-1}$; thus constituting a direct estimation of $D_{10}$ and the molecular environment in the IC 443 vicinity
under the assumed validity of this model. We have mentioned above that we assumed the $\gamma$-ray emissivity was constant within the clouds; i.e. we are assuming that there is no significant cosmic-ray gradient in the target. This assumption is an approximation, which is better when the size of the cloud is less than the distance to the accelerator and the diffusion coefficients inside and outside the cloud are not significantly different (or even if they are, the proton-proton timescale is larger than the time it takes for cosmic rays to overtake the whole cloud). 
%
%
In the case of IC 443, these conditions can be accommodated for the solutions in Table 1, except perhaps for the very massive cloud located close to the SNR at d$_{snr}=2$ pc; for which would imply a cloud average density higher than usually found, although even this might also be possible given the small scale clumps found therein (e.g., see Rosado et al. 2007).

 \begin{table}
\begin{center}
\label{tab1}
\caption{Main model parameters for solutions shown in Figure \ref{deg}. $D_{GMC}$ and $d_{snr}$ are the distance to the GMC and the closer-to-the-SNR molecular clouds. The three $f$ values quoted define $M_{GMC}=(1/f)$ 8000 M$_\odot$. The three groups explore different degeneracies: in position of the GMC, of the smaller cloud, and on the diffusion coefficient. Model 3 is not shown in Figure \ref{deg} but rather in Figure \ref{new-model}.}
\vspace{5pt}
\small
\begin{tabular}{ccccccc}
\hline \hline
Model & $D_{GMC}$ & $d_{snr}$ & $D_{10}$ & $f$  & $M_{snr}$   \\
  & pc & pc &   cm$^2$ s$^{-1}$  & \ldots &  M$_\odot$ & \\ \hline
1& 8 & 4 & 10$^{26}$ & 3.0 -- 4.0 -- 5.0 & 350\\
2& 9 & 4 & 10$^{26}$ & 1.8 -- 2.5 -- 3.2 & 350\\
3&  10 & 4 & 10$^{26}$ & 1.5 -- 2.0 -- 2.5 & 350\\
4&11 & 4 & 10$^{26}$ & 1.0 -- 1.4 -- 1.8 & 350\\
5&12 & 4 & 10$^{26}$ & 0.7 -- 1.0 -- 1.3 & 350\\
\hline
6&10 & 2 & 10$^{26}$ & 1.5 -- 2.0 -- 2.5 & 1750\\
7&10 & 3 & 10$^{26}$ & 1.5 -- 2.0 -- 2.5 & 580\\
8&10 & 5 & 10$^{26}$ & 1.5 -- 2.0 -- 2.5 & 250\\
9&10 & 6 & 10$^{26}$ & 1.5 -- 2.0 -- 2.5 & 195\\
\hline
10&10 & 4 & 8 $\times 10^{25}$ & 1.5 -- 2.0 -- 2.5 & 350\\
11&10 & 4 & 9 $\times 10^{25}$ & 1.5 -- 2.0 -- 2.5 & 350\\
12&10 & 4 & 2 $\times 10^{26}$ & 1.5 -- 2.0 -- 2.5 & 350\\
13&10 & 4 & 3 $\times 10^{26}$ & 1.5 -- 2.0 -- 2.5 & 350\\
\hline
\hline
\end{tabular}
\end{center}
\end{table}

\subsection{Influence of the $\delta$-parameter}

\begin{figure*}
\centering
\includegraphics[width=0.45\columnwidth,trim=0 5 0 10]{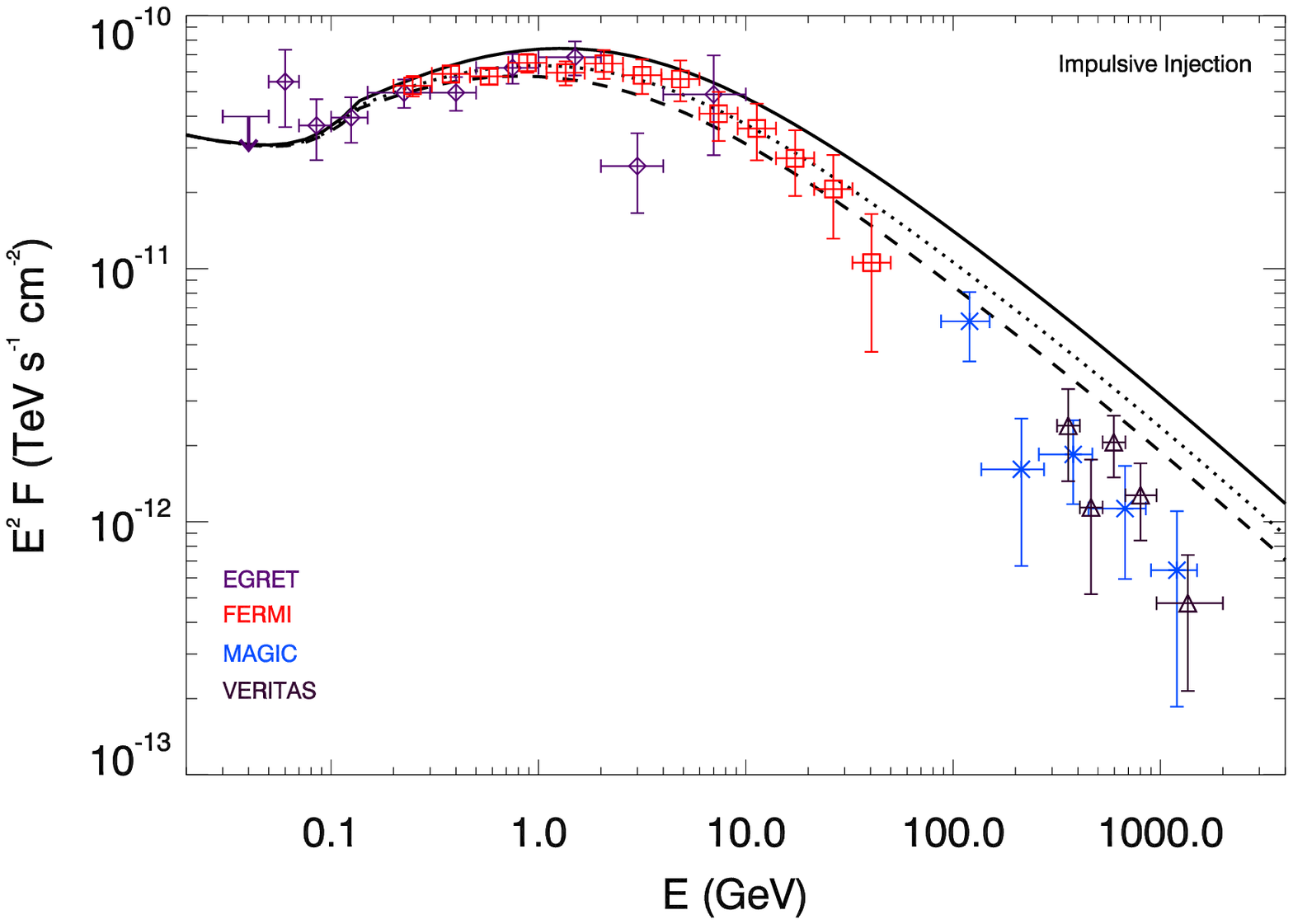}
\includegraphics[width=0.45\columnwidth,trim=0 5 0 10]{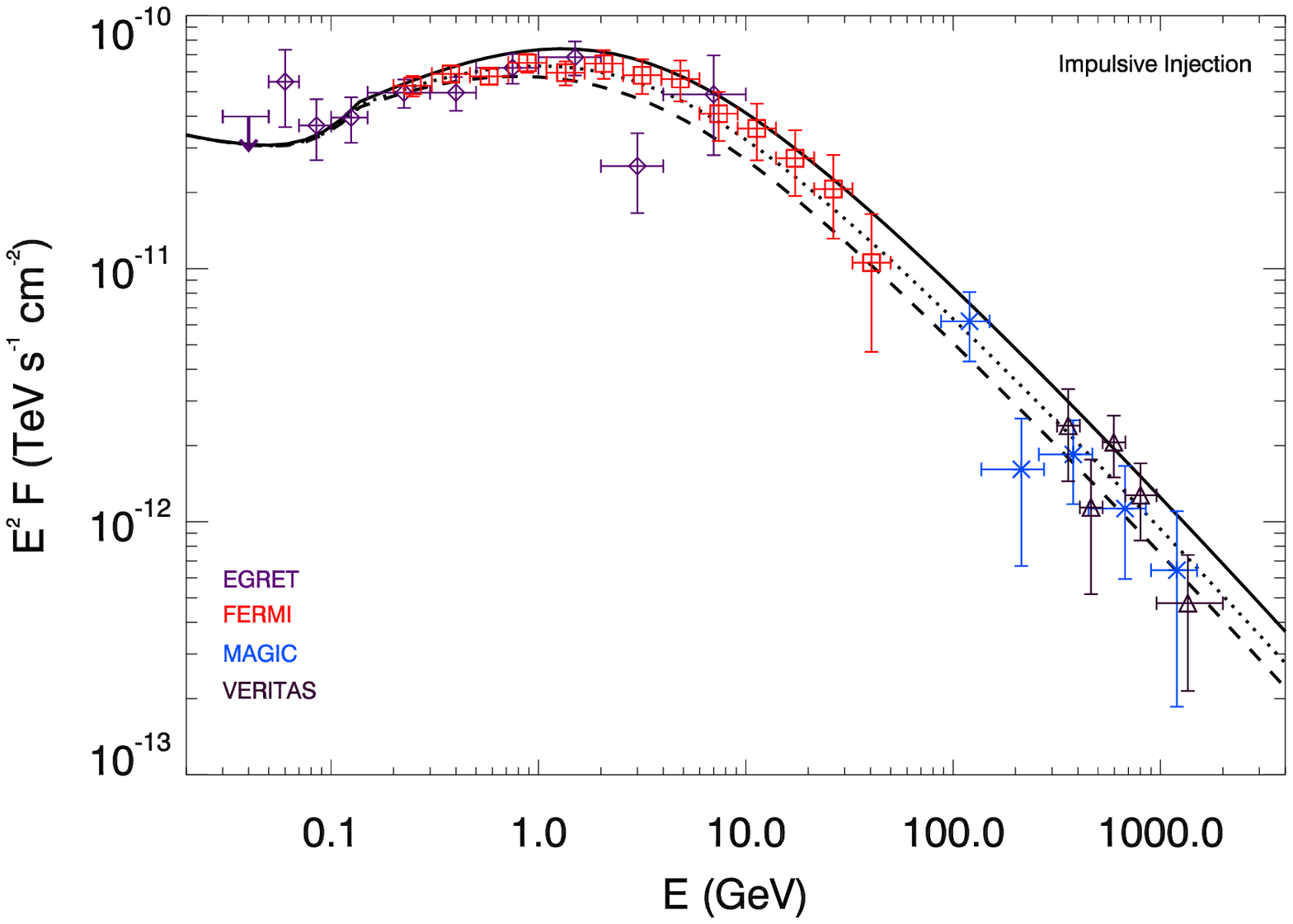}
\includegraphics[width=0.45\columnwidth,trim=0 5 0 10]{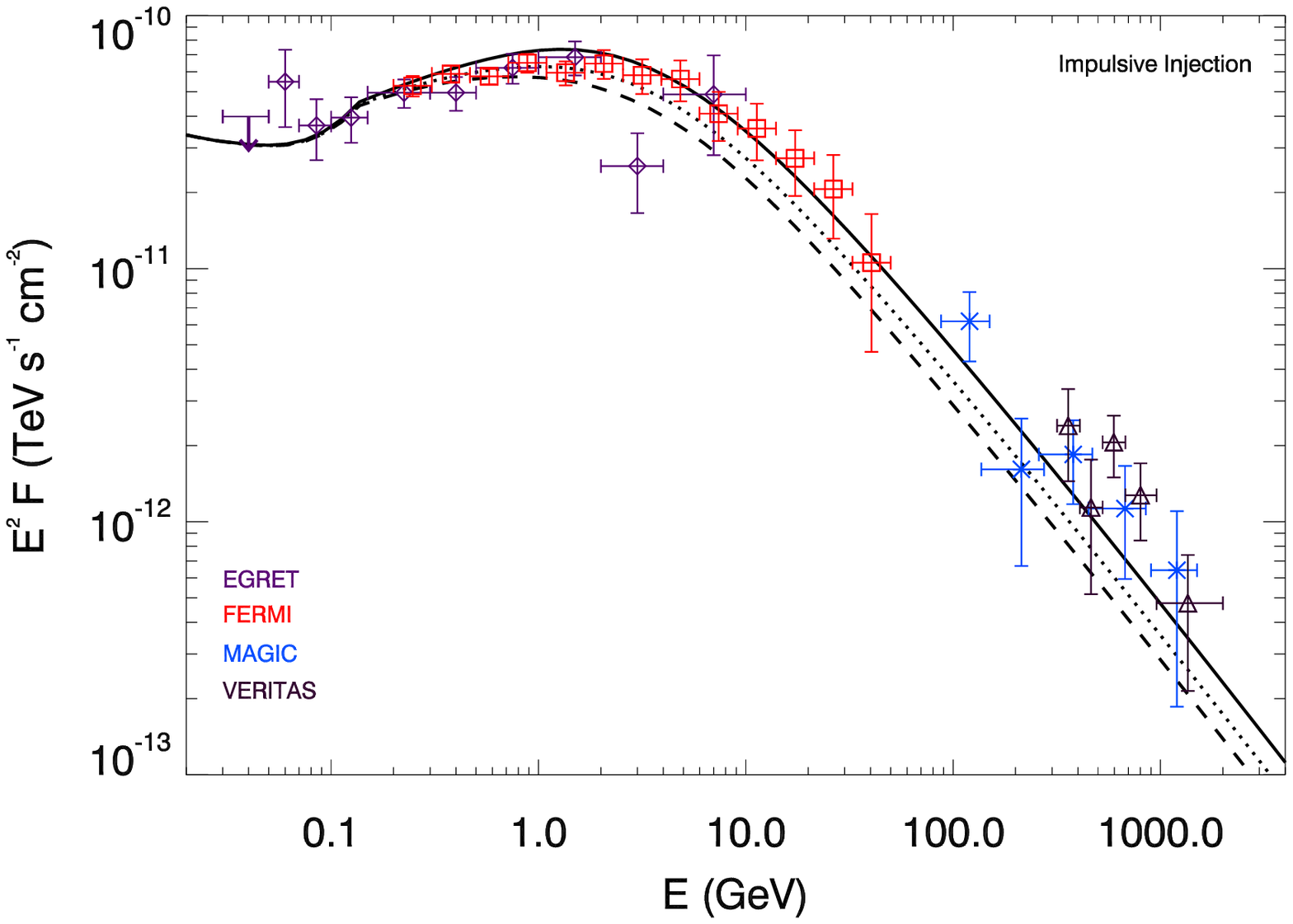}
\includegraphics[width=0.45\columnwidth,trim=0 5 0 10]{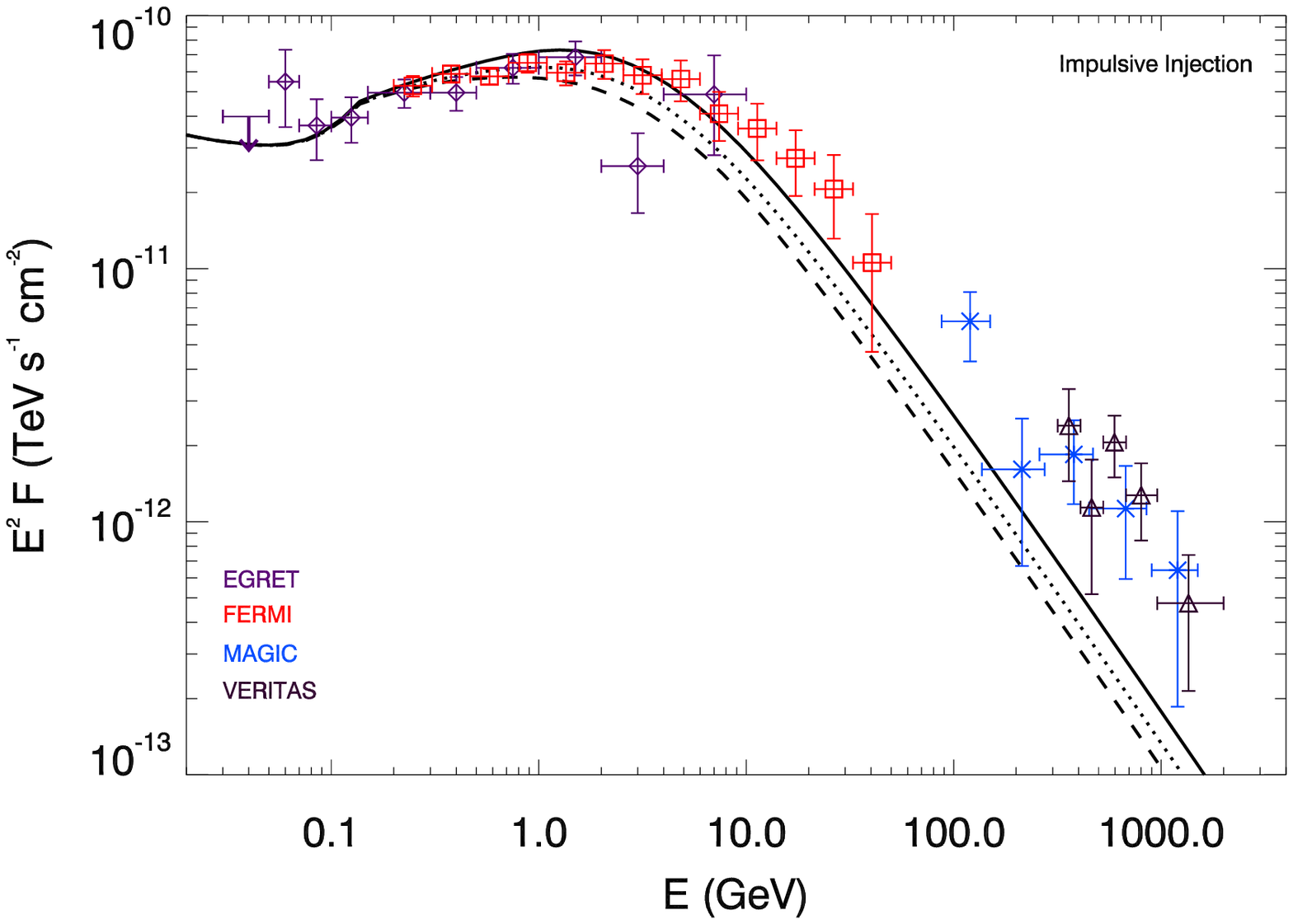}
 \caption{Comparing $\gamma$-ray yields with different $\delta$ parameters, from left to right $\delta=0.4$, 0.5, 0.6, and 0.7. Other parameters are as in Figure 2.}
\label{delta}
\end{figure*}

{ We have also made an exploration of other parameters of the model, as for instance, those influencing the way in which the diffusion coefficient varies with energy (the parameter $\delta$), or the injection spectrum of cosmic rays (referred to as $\alpha$), and came to the conclusion that their corresponding values are rather constrained. For instance, the $\delta$ parameter is expected to be around $\delta = 0.4 - 0.7$
(e.g., Berezinskii et a. 1990) and a typical value of 0.5 is usually assumed. Figure \ref{delta} gives account of how small variations in $\delta$ change the slope of  good-fitting solutions to the high and very-high energy data. One can see that for steeper $\delta$-parameters, of course, steeper $\gamma$-ray spectrum are found. If the masses of the molecular clouds are maintained and $\delta$ is larger, in order to have a good fit one would need an even lower $D_{10}$, lower than $D_{10}=10^{26}$ cm$^{2}$ s$^{-1}$ (see Figure \ref{deg}), making the solution less feasible.  }

\subsection{Uncertainties due to the cross section parameterization}

\begin{figure*}
\centering
\includegraphics[width=0.33\columnwidth,trim=0 5 0 10]{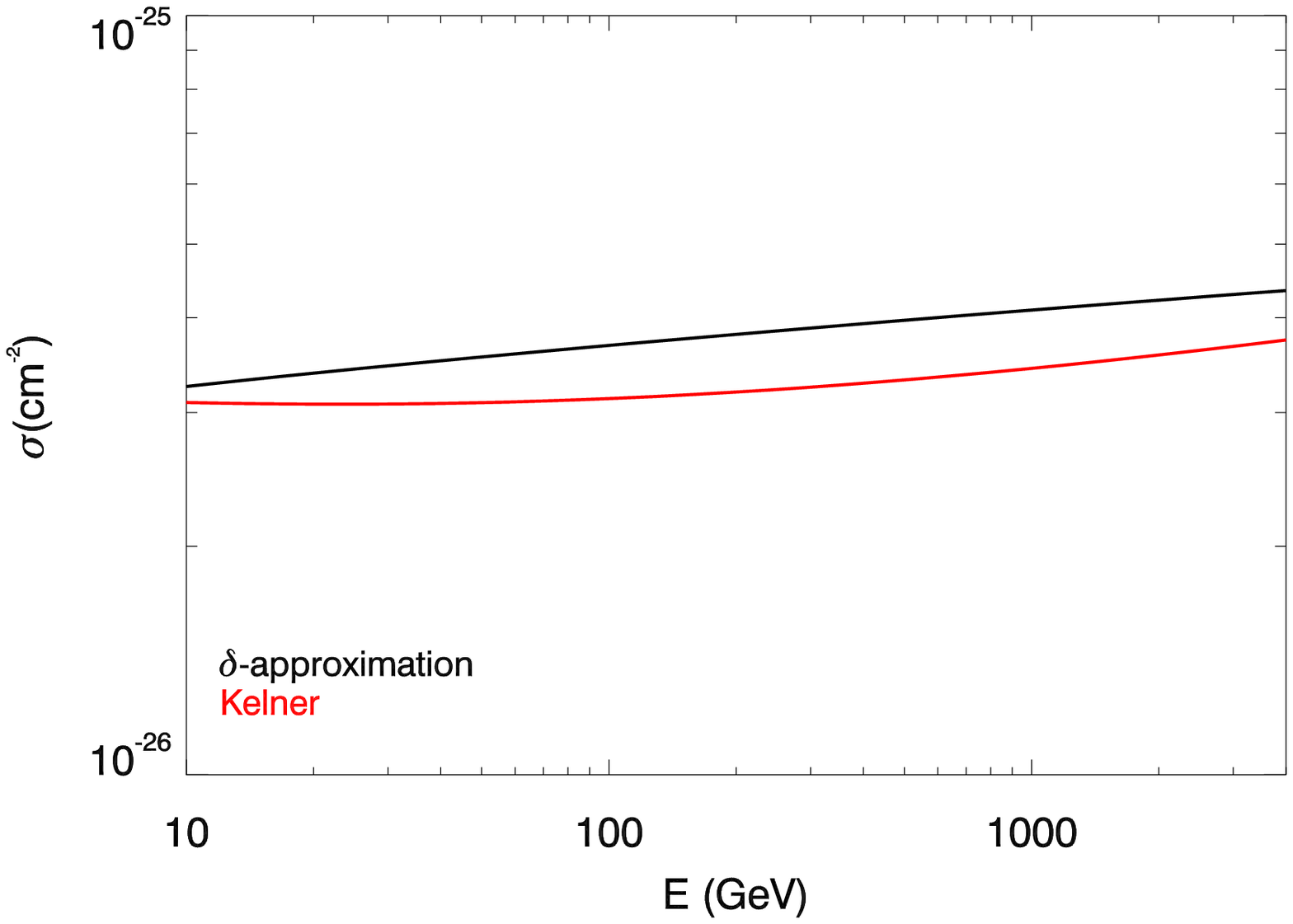}
\includegraphics[width=0.33\columnwidth,trim=0 5 0 10]{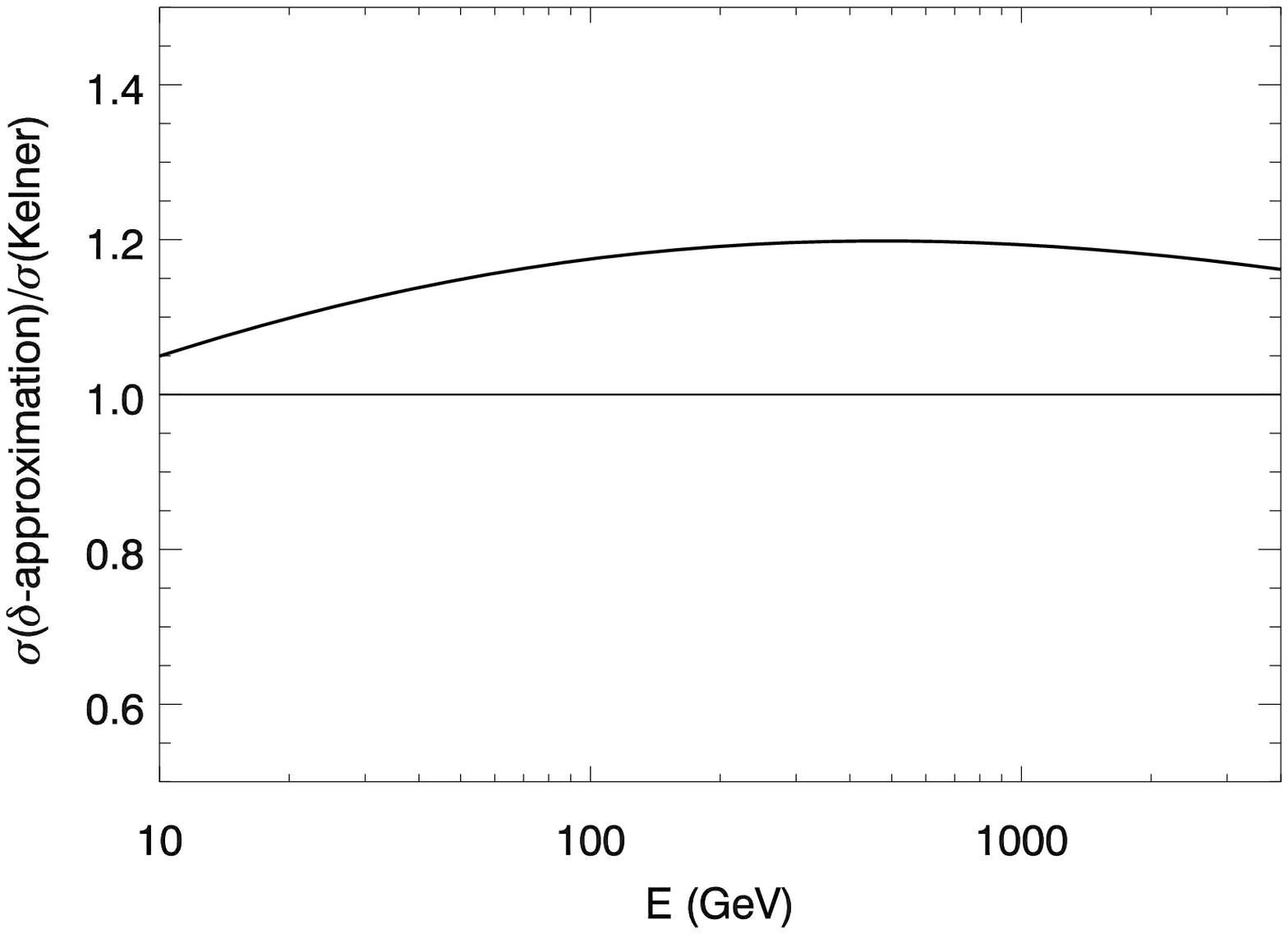}
 \caption{Comparison (left) and ratio (right) of the cross section parameterizations used in the figures above, the $\delta$-functional form by Aharonian \& Atoyan (1996),  with that of Kelner et al. (2006). }
\label{cs}
\end{figure*}

{ We have checked whether changes in the cross section parameterization can produce significant variance in the results.
In the appendix of Domingo Santamar\'ia and Torres (2005), the different predicted yields in $\gamma$-rays obtained when using alternate cross section parameterizations known by then were compared  among themselves and with data. The parameterizations therein considered were Kamae et al.'s (2005); the $\delta$-functional form by Aharonian \& Atoyan (1996) that is used above, Stephen and Badwhar's (1981), and Blattnig et al.'s (2000a,b). It was found that Kamae's and the $\delta$-functional form were very close to each other, as seen in Fig. 11 of that paper, which showed the $\gamma$-ray emissivities obtained with the corresponding use of each of the parameterizations of the cross section. In that paper, it was also found that neither Stephen and Badwhar's (1981) nor Blattnig et al.'s (2000a,b) were appropriate for their use in broad-band high-energy modeling such as the one we pursue here.
More recently, Kelner et al. (2006) presented a new approach for obtaining the cross section in $pp$ interactions. These authors used 2 shapes for representing the cross section, separated in energy. At low energies (up to 100 GeV), Kelner et al. approach uses a slightly modified but similarly-shaped $\delta$-functional form. At high-energy, its approach is different, and consists of presenting an analytical shape fitting of the results of the simulations of energy distribution of $\pi$ mesons by the SYBILL code. Figure \ref{cs} shows a comparison of the cross section parameterizations used in the figures above, the $\delta$-functional form by Aharonian \& Atoyan (1996),  with that of Kelner et al. (2006). Differences are within 20\%, with the  $\delta$-functional approximation being larger.}

\begin{figure*}
\centering
\includegraphics[width=0.45\columnwidth,trim=0 5 0 10]{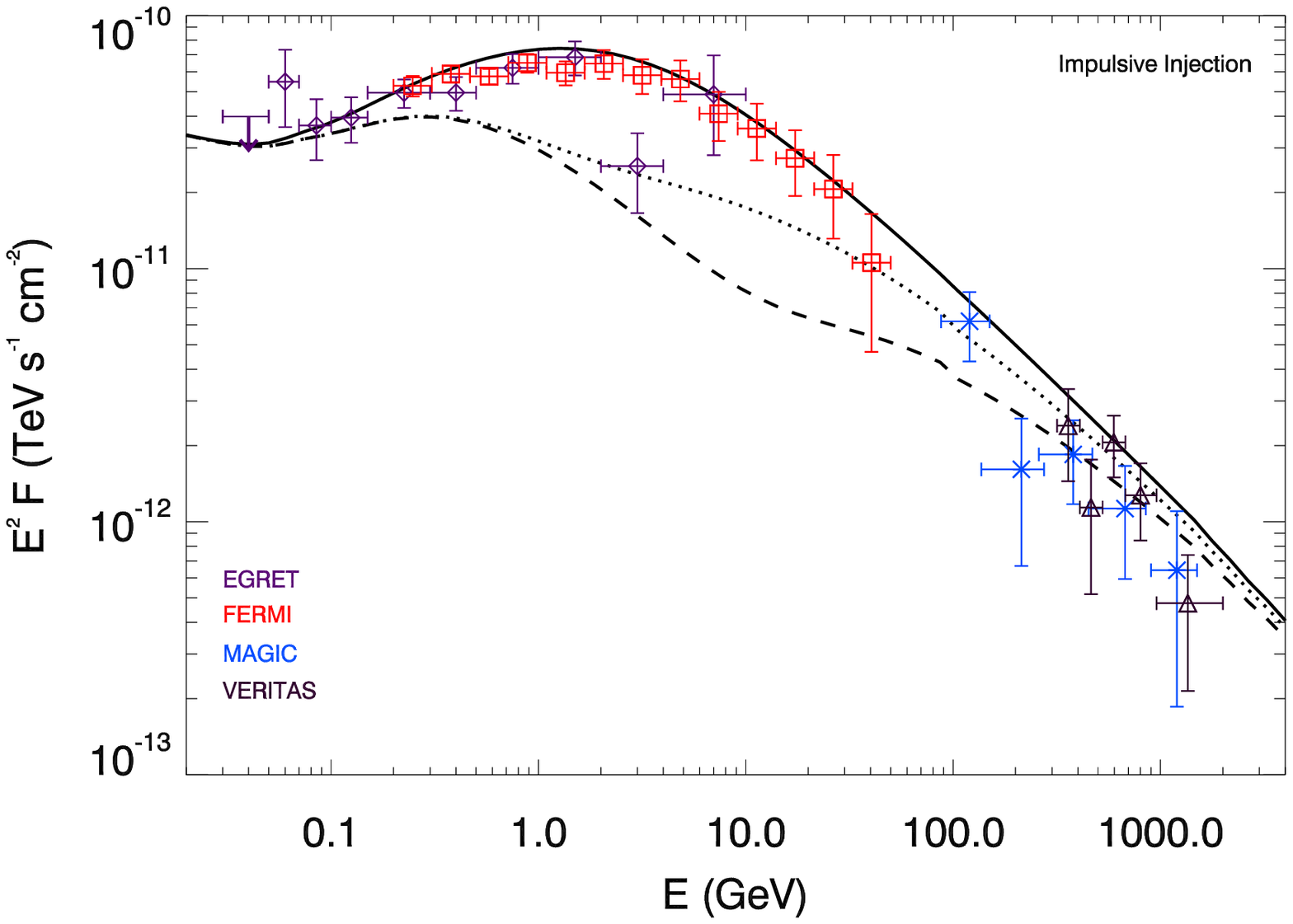}
\includegraphics[width=0.45\columnwidth,trim=0 5 0 10]{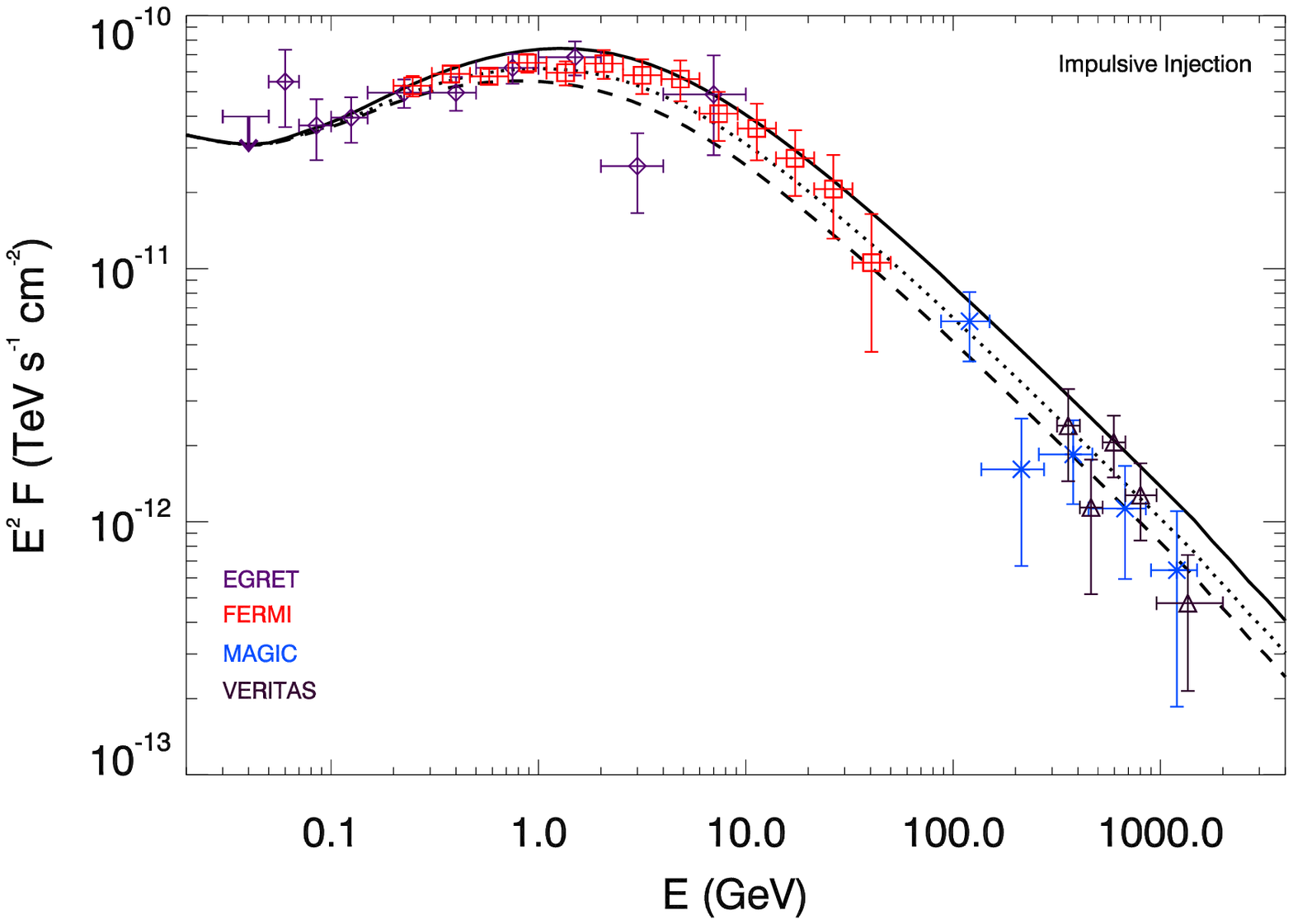}
\caption{Left: $\gamma$-ray flux results using the Kelner et al. (2006) approximation for different distances from the shell to the TeV-producing cloud, 10 (solid), 15 (dotted) and 20 (dashed) pc. The close-to-the-remnant cloud is fixed at 4 pc and contains 350 M$_\odot$ -- changes in these latter value do not improve the overall fit. Right: For the best-fitting distance models,  $\gamma$-ray flux results using the Kelner et al. (2006) for different values of the giant cloud mass (see text for details).
 }
\label{cs2}
\end{figure*}

{ The concomitant change in flux predictions, produced only because of a different use of a cross section parameterization, then, can be reabsorbed as part of the uncertainty in the determination of the model. For instance, Figure \ref{cs2} shows its impact in two ways: The left panel shows several alternatives for the position of the large (TeV-producing) cloud in the model; located at 10, 15, and 20 pc (here we maintain the mass of this cloud fixed 
and look only at the shape of the curves for different distances). 
It is clear that the change in cross section parameterization does not make any of the previously unfeasible models, feasible, and again single out a distance of about 10 pc from the SNR shell to the giant TeV-producing cloud for obtaining a good fit. The right panel assumes this distance of 10 pc and explores the uncertainty in the determination of the cloud mass. The parameters therein shown are 4 pc, and 350 M$_\odot$ for the close-to-the-remnant cloud, i.e., the same as above, and 10 pc and 7272, 5333, 4210 M$_\odot$  for the TeV-producing giant cloud. }

\subsection{Computation of secondaries other than photons}

\begin{figure*}
\centering
\includegraphics[width=0.45\columnwidth,trim=0 5 0 10]{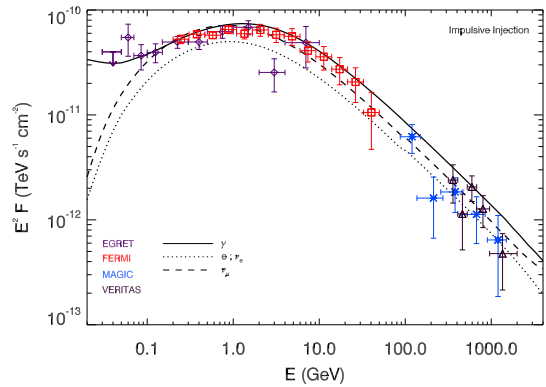}
\caption{Electrons (electrons and positrons are shown together), photons and two flavors of neutrinos produced within the clouds considered nearby IC 443, using a
set of parameters shown in Figure \ref{cs2}, right panel, with mass of the giant cloud equal to 7272 M$_\odot$.
The $\nu_\mu$ and $\nu_e$ neutrino curves show the particle and the anti-particle flux together. Data should only be compared with the photon curve.}
\label{cs3}
\end{figure*}

{ With the use of the Kelner et al. (2006) parameterization one can also readily compute secondaries other than photons, and this is shown in Figure \ref{cs3}. Gabici et al. (2009) showed that secondary electrons produced within clouds of a wide range of parameters can escape without being affected by significant losses; i.e. that the propagation time through the cloud for cosmic-ray electrons is shorter than the energy loss time for particles energies between $\sim 100$ MeV and few hundreds TeV. There would be, then, little effect of the secondary electrons produced on the non-thermal emission from the cloud. In addition, for typical densities of clouds, in the several hundreds to several thousands particles per cm$^{3}$, the dominant energy loss from $\sim$ 100 MeV and $\sim$ 10 TeV, would be bremsstrahlung and not synchrotron. }

{ A conclusive proof of the hadronic nature of the gamma-ray emission could however come from the detection of neutrinos. 
Neutrino telescopes search for up-going muons produced deep in the
Earth, and are mainly sensitive to the incoming flux of $\nu_\mu$
and ${\bar \nu}_\mu$.  The finished ICECUBE, for example,  will consist of 4800
photomultipliers, arranged on 80 strings placed at depths between
1400 and 2400 m under the South Pole ice (e.g., Halzen 2006). The
strings will be located in a regular space grid covering a surface
area of 1 km$^2$. Each string will have 60 optical modules (OM)
spaced 17 m apart. The number of OMs which have seen at least one
photon (from \v{C}erenkov radiation produced by the muon which
resulted from the interaction of the incoming $\nu$ in the earth and
ice crust) is called the channel multiplicity, $N_{\rm ch}$. The
multiplicity threshold is set to $N_{\rm ch}=10$, which corresponds
to an energy threshold of 200~GeV. The angular resolution of ICECUBE will be around
$\sim 0.7^\circ$.

A first estimation of the event rate of the atmospheric
$\nu$-background that will be detected in the search bin can be obtained as (e.g., Anchordoqui et al. 2003)
\begin{equation}
\left. \frac{dN}{dt}\right|_{\rm B} = A_{\rm eff}\, \int dE_\nu
\,\frac{d\Phi_{\rm B}}{dE_\nu}\, P_{\nu \to \mu}(E_\nu)\,\,\Delta
\Omega\,, \label{background}
\end{equation}
where $A_{\rm eff}$ is the effective area of the detector, $\Delta
\Omega \approx 1.5 \times 10^{-4}$~sr is the angular size of the
search bin, and $d\Phi_{\rm B}/dE_\nu \lesssim 0.2 \,(E_\nu/{\rm
GeV})^{-3.21}$~GeV$^{-1}$ cm$^{-2}$ s$^{-1}$ sr$^{-1}$  is the
$\nu_\mu + \bar \nu_\mu$ atmospheric $\nu$-flux~(Volkova 1980,
Lipari 1993). Here, $P_{\nu \to \mu} (E_\nu)$ denotes the
probability that a $\nu$ of energy $E_\nu$ on a trajectory through
the detector, produces a muon.  For $E_\nu \sim 1 - 10^3~{\rm
GeV}$, this probability is $ \approx 3.3 \times 10^{-13}
\,(E_\nu/{\rm GeV)}^{2.2}$, whereas for $E_\nu
> 1~{\rm TeV}$,
$P_{\nu \to \mu} (E_\nu) \approx 1.3 \times 10^{-6} \,(E_\nu/{\rm
TeV)}^{0.8}$ ~(Gaisser et al. 1995).
On the other hand, the $\nu$-signal is similarly obtained as
\begin{equation}
\left. \frac{dN}{dt}\right|_{\rm S} =  A_{\rm eff} \,\int dE_\nu
\,  ( F_{\nu_\mu }+F_{{\bar \nu}_\mu} ) \,P_{\nu \to \mu}(E_\nu)\
\,, \label{yellowsubmarine}
\end{equation}
where $( F_{\nu_\mu }+F_{{\bar \nu}_\mu} )$ is the incoming
$\nu_\mu$-flux.
In the previous integrals we use both expressions for $P_{\nu \to \mu} (E_\nu)$
according to the energy, and integrate from 200 GeV up to 10 TeV. 
The effect of $\nu$-oscillations is taken into account following table 2 of Cavassini et al. (2006), where 
the oscillation probability in the average vacuum oscillation hypothesis is given. It is is assumed that the inter-conversion probability
between flavors and between anti-flavors is the same.
As an effect of oscillations, the flavor composition of all the expected fluxes for each flavor are within 50\% of each other.
Using the former formulae and the secondary computation shown in Figure \ref{cs3} we find that the 
number of muon neutrino signal events is 0.6 per year of observation, still significantly below than the estimation of the number of background events, which under the previous provisions is 6.4 along the same period,
with the full ICECUBE array. If we consider only events above 1 TeV, 
the expected signal is 0.25 year$^{-1}$, and the computed background is 1.92 year$^{-1}$. ICECUBE does not seem to be able to distinguish this signal in reasonable integration times, at least within the reach of  this simplified treatment of the detector. 

}

\section{Concluding remarks}

{ The recent observations of the IC 443 environment made by AGILE, {\it   Fermi}, and VERITAS
at the GeV and TeV energies are spectrally
consistent with the interpretation of cosmic-ray interactions with a giant molecular cloud lying in front of the remnant. This scenario would be producing no significant counterpart at lower energies at that spot, and would then be leading to a natural interpretation of the dislocation between the centroids of the detections at the different energy bands. Use of the latest data allowed to estimate, within the assumed validity and framework of this model, the diffusion characteristics in this environment, showing that the diffusion coefficient is lower; the cosmic-ray density is higher, than the Earth-values of these magnitudes. Uncertainties in amount and localization of target molecular mass still remains as  does also in the density at which this molecular material is found (e.g., the uncertainty in the cosmic-ray-overtaken mass discussed above is about 100\% for matching models at the extremes of this parameter). But even allowing for a range this large, the model could accommodate some but not all variations in other parameters, with the values of $D_{10}$ and distances from the SNR shell seemingly being solid constraints. } \\


\subsection*{Acknowledgments}

{\it This work has been supported by grants AYA2009-07391 and SGR2009-811. The work of E. de Cea del Pozo has been made under the auspice of a FPI Fellowship, grant BES-2007-15131. An anonymous referee is acknowledged for comments that led to an improvement of the paper. }

\label{lastpage}
\end{document}